\documentclass[showpacs,prd,superscriptaddress,onecolumn]{revtex4}

\usepackage{amsmath}
\usepackage{amssymb}
\usepackage{amsfonts}
\usepackage{graphicx,bm}
\usepackage{dcolumn}
\usepackage[colorlinks=true]{hyperref}
\usepackage{color,amsxtra}
\usepackage{epsf}
\usepackage{enumerate}
\usepackage{hhline}
\usepackage{array}
\usepackage{tabularx}
\newcommand{\be}{\begin{equation}}
\newcommand{\ee}{\end{equation}}
\newcommand{\bea}{\begin{eqnarray}}
\newcommand{\eea}{\end{eqnarray}}
\newcommand{\beaa}{\begin{eqnarray*}}
\newcommand{\eeaa}{\end{eqnarray*}}

\newcommand{\nn}{\nonumber \\}
\newcommand{\e}{\mathrm{e}}

\begin{document}

\tolerance=5000

\title{Modelling and testing the equation of state for (Early) dark energy}
\author{Shin'ichi~Nojiri}
\email{nojiri@gravity.phys.nagoya-u.ac.jp} \affiliation{Department of Physics, Nagoya University, Nagoya 464-8602, Japan}
 \affiliation{Kobayashi-Maskawa Institute for the Origin of Particles and the Universe,
 Nagoya University, Nagoya 464-8602, Japan}
\author{Sergei~D.~Odintsov}
\email{odintsov@ice.csic.es} \affiliation{Institut de Ci\`{e}ncies de l'Espai,
ICE/CSIC-IEEC, Campus UAB, Carrer de Can Magrans s/n, 08193 Bellaterra (Barcelona),
Spain}
 \affiliation{Instituci\'o Catalana de Recerca i Estudis Avan\c{c}ats (ICREA),
Passeig Luis Companys, 23, 08010 Barcelona, Spain}
\author{Diego~S\'aez-Chill\'on~G\'omez}
\email{diego.saez@uva.es} \affiliation{Department of Theoretical, Atomic and Optical
Physics, Campus Miguel Delibes, \\ University of Valladolid UVA, Paseo Bel\'en, 7, 47011
Valladolid, Spain}
\author{German~S.~Sharov}
 \email{sharov.gs@tversu.ru}
 \affiliation{Tver state university, Sadovyj per. 35, 170002 Tver, Russia}
 \affiliation{International Laboratory for Theoretical Cosmology,
Tomsk State University of Control Systems and Radioelectronics (TUSUR), 634050 Tomsk,
Russia}

\begin{abstract}
A general equation of state is considered for analysing the possible behaviors for (Early) dark energy that alleviates
the Hubble parameter tension problem.
By departing from the possible evolution for the (Early) dark energy density and the corresponding dynamical equations,
the equation of state is obtained, which allow us to analyze qualitatively the cosmological evolution and
the dominance of each term in the equation of state along the cosmic expansion, which show some interesting
consequences as the occurrence of (past) future singularities.
Then, by considering two general models, their free parameters are fit with different sources of data,
showing the goodness of the fits in comparison to more standard models.
Results might be considered as a promising starting point to get a better understanding of
the cosmological evolution as a whole.

\end{abstract}
%
%
\maketitle
%

\section{Introduction}

Cosmology has experienced a tremendous growth over the last decades, when cosmology has turned
from an almost only theoretical framework to a precise quantitative science that can be tested with real observational data.
This has also led to the discovery of unexplained phenomenas that are entailing considerable challenges
both theoretically as observationally.
Currently one of the main aims in cosmology is focusing on the explanation of the late-time acceleration
of the universe expansion, which is widely supported by different observations and which requires the assumption
of an effectively perfect fluid that violates some of the energy conditions, the so-called dark energy
(for a review see \cite{Bamba:2012cp,Huterer:2017buf}).
The research on dark energy has involved a great effort from the scientific community that has given rise not only
to a large number of proposals to try to explain the late-time acceleration phenomena but also
to new theoretical frameworks that has open new lines of research.
Nevertheless, despite the large number of viable theoretical models that fit well the data,
there is no convincing reason to believe that the problem is close to being solved,
although the well-known $\Lambda$CDM model is assumed to be at least the departure for understanding
the problem and its possible solution, as it is the simplest one and involves a reduced number of free parameters.
In addition, over the last years the tension among the values of the Hubble constant as estimated
from different sources of data has implied another challenge in cosmology to find a cause for such discrepancy.
Basically, the measurements from the Cosmic Microwave Background (CMB), ``early'' universe,
by Planck collaboration \cite{Aghanim:2018eyx} reveal a significant difference
with respect to those estimations realised from other sources that collect data from late universe,
particularly by the SHOES collaboration of the Hubble Space Telescope \cite{Riess:2019cxk}
that includes data from Type Ia Supernovae (SNe Ia), exceeding in this case $4\sigma$
with respect to the estimation by Planck (for a review on the Hubble tension and
the different discrepancies see \cite{Verde:2019ivm,DiValentino:2021izs} and for a summary on the possible solutions see \cite{DiValentino:2020zio}).
The solution to this tension  might lie on systematics errors in data, but this will imply
the same amount of systematic errors in different data from independent sources,
such that seems dismissed.
Other approaches for alleviating such tension include the tip of the red giant branch (TRGB)
by the Carnegie-Chicago Hubble Program (CCHP) \cite{Freedman:2020dne},
independent measurements of lensed quasars and time delays by H0LiCOW project \cite{Wong:2019kwg}, megamaser hosting galaxies \cite{Pesce:2020xfe}, dividing the Pantheon sample in bins \cite{Dainotti:2021pqg} and by using independent geometrical datasets \cite{Renzi:2020fnx}, among others.
Nevertheless, the problem might be approached from a more theoretical framework,
as the estimation of the Hubble constant in early universe relies on angular scales,
which depend on the ratio of the physical scale and the angular distance to the CMB,
which might be modified in such a way to provide a different estimation of the Hubble constant.
To do so, an additional energy component is required to become important just in a narrow period
of the expansion history after the matter-radiation equality, which has been called early dark energy,
and which might be a new field \cite{Poulin:2018cxd,Mortsell:2018mfj,Niedermann:2020dwg,Garcia:2020sjl,Ye:2020btb}, interactions
among dark energy and dark matter
\cite{DiValentino:2017iww,Yang:2018euj,Pan:2019jqh,Gomez-Valent:2020mqn,Pan:2020zza},
relativistic species
\cite{DEramo:2018vss,Vagnozzi:2019ezj} or modifications of general relativity \cite{Nunes:2018xbm,Wang:2020zfv,Odintsov:2020qzd,Braglia:2020auw}.

A useful and well-known analysis when dealing with dark energy lies on describing it as a fluid
with an effective equation of state (EoS), which do not infer about the theoretical origin of dark energy
but presents a useful approach to know better about the behavior and form of its EoS
and consequently on its evolution along the universe expansion to be fit with observational data
and then compared to other models.
This type of approach arose soon after the  discovery of late-time acceleration, being a useful way
for ruling out models and for testing the strength of $\Lambda$CDM model \cite{Huterer:2000mj}.
Over the years, such parametrizations of the EoS for dark energy have become more complex
and sophisticated, accounting for different behaviors at different epochs for dark energy
and providing a way to reconstruct the underlying theoretical description for dark energy
(for a review see \cite{Bamba:2012cp}). In this sense, a general equation of state was proposed
in \cite{Nojiri:2005sr,Capozziello:2005pa}, which is reconstructed from the cosmological evolution by departing from the FLRW equations. In this sense, the EoS for dark energy can be also unveiled by using different observational datasets \cite{Gerardi:2019obr}. We should note that a particular class of such imperfect fluids is a viscous fluid,
such that dark energy with bulk viscosity has been widely analyzed in the literature
\cite{Brevik:2017msy,Cataldo:2005qh,Cruz:2018yrr}.
Moreover, some particular parametrizations can account for transitions in the universe evolution, as the case
for phantom dark energy \cite{Elizalde:2008yf,Leanizbarrutia:2014xta}, whereas others can be related as the manifestation
of some modification of the general relativity \cite{SaezGomez:2008uj}.

In the present paper, we generalize the above works by constructing the corresponding EoS in terms
of its dependence on the scale factor and the Hubble parameter.
For any perfect fluid with a constant EoS, the continuity equation forces the energy density to behave
as a power-law of the scale factor, then a natural choice suggests that dark energy dependence will go also
as a power of the scale factor at least at some limit, a natural choice suggests that dark energy dependence
will go also as a power of the scale factor, at least at some limit, but likely more complex with different terms
and even depending on the Hubble parameter.
Moreover, an explicit dependence on the scale factor $a$ of the universe might have some analogies
with a superfluid EoS \cite{Volovik:2000ua,Rosu:2020tov}. In addition, the inclusion of different terms provides
a way for describing dark energy and early dark energy by the same EoS.
Hence, we present a way of reconstructing such EoS and study the qualitative behavior
of the cosmological evolution depending on the term that dominates and the possible occurrence
of future and past singularities.
Then, by using different observational data sets, two general EoS are fit and confronted to $w$CDM, a generalisation of $\Lambda$CDM,
providing an interesting and new way for reconstructing effectively the EoS for (early) dark energy.

The paper is organized as follows: in Section~\ref{Sec2}, we introduce the general EoS and its reconstruction
in terms of the FLRW equations. Section~ \ref{Sec3} is devoted to a simple toy model and the qualitative analysis
of the cosmological expansion.
In Section~\ref{Sec4}, we analyze some more complex EoS and the behaviors of the expansion in the early
and late universe.
In Section~\ref{Sec5} several models for early dark energy are proposed in terms of its EoS,
we also introduce the two main models of this paper.
Section~\ref{Sec6} refers to the observational datasets used in the paper to confront those two models.
The results of the fittings of the models to the data are covered in Section~\ref{Sec7}.
Finally, Section~\ref{Sec8} gathers the summary and conclusions of the paper.

\section{Generalising the Equation of State \label{Sec2}}

Let us start by considering the following general equation of state \cite{Nojiri:2005sr},
\begin{equation}
\label{GEoS0}
p= - \rho + h\left( \rho, a, H, \dot H, \cdots \right) \, .
\end{equation}
Here $p$ is the pressure, $\rho$ is the energy density, and
$h\left( \rho, a, H, \dot H, \cdots \right)$ is a function of $\rho$, the scale factor $a$,
the Hubble rate $H=\dot a/a$, $\dot H$, and so on.
The corresponding FLRW equations are given by
\begin{equation}
\label{GEoS1B}
\frac{3}{\kappa^2} H^2 = \rho\, , \quad  - \frac{1}{\kappa^2} \left( 2 \dot H + 3H^2 \right)
= p \, .
\end{equation}
As far as $p$ and $\rho$ express the total pressure and energy density, by using the first and
second FLRW equations~(\ref{GEoS1B}), the EoS~(\ref{GEoS0}) can be written as
\begin{equation}
\label{GEoS0A}
p= - \rho + h\left( \rho, a, \sqrt{ \frac{\kappa^2}{3}\rho},
- \frac{\kappa^2}{2} \left( p + \rho \right), \cdots \right) \, .
\end{equation}
Hence, the EoS can be rewritten by using $p$, $\rho$, $a$, and possibly,
$\ddot H$, $\dddot H$, etc. This expression represents the most general one that can be written
for a perfect fluid in the FLRW spacetime. Nevertheless, one might consider simpler cases than
this generalization. As an example, we consider the following energy density evolution,
\begin{equation}
\label{GEoS1}
\rho = f\left( H, a\right)\, .
\end{equation}
Then, by using the continuity equation,
\begin{equation}
\label{EDE2BB}
0 = \dot \rho + 3 H \left( \rho + p  \right) \, .
\end{equation}
The corresponding EoS yields,
\begin{equation}
\label{GEoS3}
\frac{p}{\rho}
= -1 - \frac{1}{3 f\left( H, a\right)} \left( \frac{\dot H}{H} \partial_H f\left( H, a\right)
+ a\partial_a f\left( H, a\right) \right) \, .
\end{equation}
We can use again the FLRW equations (\ref{GEoS1B}) to express the EoS (\ref{GEoS3}) just
in terms of the pressure, the energy density and the scale factor as
\begin{equation}
\label{GEoS3B}
\frac{p}{\rho}
= -1 - \frac{1}{3 f\left( \sqrt{ \frac{\kappa^2}{3}\rho}, a\right)}
\left( \frac{ - \frac{\kappa^2}{2} \left( p + \rho \right)}{\sqrt{ \frac{\kappa^2}{3}\rho}}
\left. \left(\partial_H f\left( H, a\right) \right) \right|_{H=\sqrt{ \frac{\kappa^2}{3}\rho}}
+ \left. a\left( \partial_a f\left( H, a\right) \right) \right|_{H=\sqrt{ \frac{\kappa^2}{3}\rho}} \right) \, .
\end{equation}
For instance, for a perfect fluid with the following dependence on the scale factor,
\begin{equation}
\rho\propto a^{m}\, .
\label{Ex1}
\end{equation}
The corresponding EoS (\ref{GEoS3B}) leads to
\begin{equation}
\frac{p}{\rho}=-1-\frac{1}{3} m\, ,
\end{equation}
which can be easily identified with a perfect fluid with constant EoS $p/\rho=w$ by setting
$m=-3(1+w)$. In the same way, we might consider
\begin{equation}
\rho\propto H^n\, ,
\end{equation}
and we find that the corresponding EoS leads to
\begin{equation}
\label{GEoS7}
\frac{p}{\rho} = -1 - \frac{n\dot H}{3 H^2} \, .
\end{equation}
where for $n=0$, the EoS is the one of a cosmological constant.

Hence, by specifying the corresponding dependence of the energy density,
one can easily reconstruct the EoS by (\ref{GEoS3}) and (\ref{GEoS3B}).
In the next sections, we will consider more complex cases that might describe several perfect fluids. \\

Note that the above EoS (\ref{GEoS3}) can be easily realised in multiple frameworks, by the appropriate scalar field Lagrangian \cite{Elizalde:2008yf} or through modifications of GR, as in the case of $f(R)$ gravities \cite{SaezGomez:2008uj}, among others. Let us consider the former to show the reconstruction by assuming the following action:
\be
S=\int d^4x\sqrt{-g} \left[\frac{1}{2\kappa^2}R-\frac{1}{2} \omega(\phi)\partial_{\mu}\phi\partial^{\mu}\phi-V(\phi)\right]+S_m\ .
\label{actionSF}
\ee
Here, $S_m$ accounts for the matter content. Then, the FLRW equations lead to:
\be
\frac{3}{\kappa^2}H^2=\rho=\rho_m+\rho_{\phi}\ , \quad -\frac{2}{\kappa^2}\dot{H}=\rho+p=\rho_m(1+w_{m})+\rho_{\phi}+p_{\phi}\ ,
\label{FLRWeqsSF}
\ee
where:
\be
\rho_{\phi}=\frac{1}{2}\omega(\phi)\dot{\phi}^2+V(\phi)\ , \quad p_{\phi}=\frac{1}{2}\omega(\phi)\dot{\phi}^2-V(\phi)
\ee
For a given solution $a=a(t)$, we can redefine the scalar field such that $\phi=t$, and the Hubble parameter and the scale factor can be expressed as functions of the scalar field $\phi$. By the FLRW equations (\ref{FLRWeqsSF}), the corresponding kinetic term $\omega(\phi)$ and the potential $V(\phi)$ for the scalar field are obtained:
\bea
\omega(\phi)&=&-\frac{H^2}{\kappa^2 f(H,a)}\left( \frac{\dot H}{H} \partial_H f\left( H, a\right)
+ a\partial_a f\left( H, a\right) \right)-(1+w_m)\rho_0 a^{-3(1+w_m)}\ , \nn
V(\phi)&=&\frac{3}{\kappa^2}H^2+\frac{H^2}{2\kappa^2 f(H,a)}\left( \frac{\dot H}{H} \partial_H f\left( H, a\right)
+ a\partial_a f\left( H, a\right) \right)+\frac{1}{2}(w_m-1)\rho_0 a^{-3(1+w_m)}\ .
\eea
Hence, the corresponding action for the scalar field is reconstructed departing from the generic EoS (\ref{GEoS3}). Here we will focus just on the phenomenological description of the EoS and its behaviour regardless of the underlying theory that might reproduce such EoS.

\section{A simple model \label{Sec3}}

As a more general case than (\ref{GEoS1}), we might consider
\begin{equation}
\label{GEoS4}
f\left( H, a\right) = f^{(1)}\left( H \right) + f^{(2)} \left( a \right) \, , \quad
f^{(1)}\left( H \right) \equiv \sum_i \alpha_i H^{n_i} \, , \quad
f^{(2)}\left( a \right) \equiv \sum_i \beta_i a^{m_i} \, ,
\end{equation}
where $\alpha_i$, $\beta_i$, $n_i$, and $m_i$ are constants. As shown above,
for a perfect fluid with constant EoS parameter $w$, the energy density behaves as
$\rho_w \propto a^{-3\left( 1 + w \right)}$, so that any constribution from
such a type of perfect fluids can be included in (\ref{GEoS4}) in $f^{(2)}\left( a \right)$ with
$w=- 1 - \frac{m_i}{3}$, which also requires $\beta_i \geq 0$.
On the other hand, if the whole function $f\left( H, a\right)$ is the contribution from a unique fluid,
no particular assumptions on $\beta_i$ and $\alpha_i$ have to be imposed
as far as $f\left( H, a\right)$ is positive.

In addition, for a given scale factor $a=a(t)$, we may define
\begin{equation}
A(t)\equiv \frac{3}{\kappa^2} H^2 - f^{(1)}\left( H \right)\, ,
\end{equation}
which is a function of the cosmological time $t$.
Then, by solving the equation $a=a(t)$ with respect to $t$ as $t=t(a)$, we find
\begin{equation}
\label{GEoS5}
f^{(2)} \left( a \right) = A \left( t \left(a \right) \right) \, .
\end{equation}
Hence, an arbitrary time function of the scale factor $a=a(t)$ can be realized
by the model (\ref{GEoS4}).


In order to illustrate this procedure and the posterior analysis that facilitates
a qualitative description of the cosmological evolution, let us consider a simple case,
\begin{equation}
\label{GEoS6}
f^{(1)}\left( H \right)= \alpha H^n \, , \quad
f^{(2)}\left( a \right) = \sum_{i=1}^N \beta_i a^{m_i} \, ,
\end{equation}
As shown above, the second function in (\ref{GEoS6}) can be seen as the sum of contributions
from several perfect fluids with a constant EoS parameter $w=- 1 - \frac{m_i}{3}$.
 From now on, we are assuming $m_1< m_2 < \cdots < m_N$. In addition,
by assuming a perfect fluid whose energy density is given by $\rho^{(1)}=f^{(1)}\left( H \right)$,
the corresponding EoS is given by (\ref{GEoS7}), with $\alpha$ being positive and $n\neq 0, 2$,
since $n=2$ just leads to a redefinition of the coupling constant $\kappa^2$ and $n=0$
corredesponds to a cosmological constant, as shown in the previous section.

Let us now define the following function of the Hubble parameter that will turn
out fundamental in our analysis below,
\begin{equation}
\label{GEoS8}
B(H) = \frac{3}{\kappa^2} H^2  - \alpha H^n \, .
\end{equation}
We now assume $\alpha$ is positive.
As far as $n$ is positive, the algebraic equation $B(H)=0$
has a trivial solution $H=0$ and a non-trivial one given by
\begin{equation}
\label{GEoS9}
H=H_0 \equiv\left(  \frac{\alpha \kappa^2}{3} \right)^{\frac{1}{2-n}} \, .
\end{equation}
For negative $n$, the equation $B(H)=0$ has only one real solution (\ref{GEoS9}).
We should also note that when $n>2$, $B(H)$ has a maximum at $H=H_M$ defined by
\begin{equation}
\label{GEoS10}
H=H_M \equiv \left(  \frac{n \alpha \kappa^2}{6} \right)^{\frac{1}{2-n}} \, , \quad
B\left( H_M \right) = \frac{3\left( n - 2 \right)}{n\kappa^2}
\left(  \frac{n \alpha \kappa^2}{6} \right)^{\frac{2}{2-n}} \, .
\end{equation}
In case $0<n<2$, $B(H)$ has a minimum where $B< 0$ at $H=H_{M}$ ($0<H_M<H_0$)
and consequently $B(H)$ is a monotonically increasing function for $H>H_M$ and goes to
$+\infty$ as $H\to +\infty$.
In case $n>2$, $B(H)$ is a monotonically decreasing function
and goes to $-\infty$ as $H\to +\infty$. Even in this case, $H_0$ gives a possible minimum
of the possible Hubble rate $H$.

The function $B(H)$ is nothing but the first FLRW equation (\ref{GEoS1B}) with the energy
density given as in (\ref{GEoS4}) with (\ref{GEoS1}) and (\ref{GEoS6}).
Then, we can analyze the cosmological evolution by studying the function $B(H)$, which
through the first FLRW equation (\ref{GEoS1B}) in combination with (\ref{GEoS6}),
and by considering just one contribution for $f^{(2)}$, leads to
\begin{equation}
\label{GEoS11}
B(H) \sim \beta_N a^{m_N}\, .
\end{equation}
Firstly we aim to analyze the cosmological evolution in late times, where we assume
$a$ to be large. Then, depending on the value of $m_N$, we can distinguish the following cases:
\begin{itemize}
\item For $m_N<0$, the l.h.s. in (\ref{GEoS11}) decreases with the expansion
and therefore the l.h.s. can be neglected as the universe asymptotically goes to
a de Sitter Universe where $H$ is given by $H_0$ in (\ref{GEoS9}).
\item For $m_N>0$, the energy density corresponds to a phantom fluid.
Because $B(H)$ can be larger and larger if $n<2$, a Big Rip singularity occurs
in the future. On the other hand, whether $n>2$, $B(H)$ has a maximum given
in (\ref{GEoS10}) and consequently
by the Eq.~(\ref{GEoS11}), we have a maximum $a_M$ for the scale factor,
\begin{equation}
\label{GEoS12B}
a_M = \left( \frac{B\left( H_M \right)}{\beta_N} \right)^{\frac{1}{m_N}}
= \left( \frac{1}{\beta_N} \left(  \frac{n \alpha \kappa^2}{6} \right)^{\frac{2}{2-n}}
\right)^{\frac{1}{m_N}} \, .
\end{equation}
In order to study the behavior around $a\sim a_M$, the scale factor can be expressed as follows,
\begin{equation}
\label{GEoS12}
a= a_M \e^{n(t)}\, .
\end{equation}
Here $n(t)$ is negative and we can assume that $\left| n(t) \right|\ll 1$ around $a\sim a_M$.
As the Hubble parameter is given by $H=\frac{dn}{dt}$, we may assume
\begin{equation}
\label{GEoS13}
n=H_M \left( t - t_M \right) + \delta N \, .
\end{equation}
And $H=H_M$ for $t=t_M$. By expanding the function $B(H)$ around $t=t_M$,
the equation (\ref{GEoS11}) turns out
\begin{equation}
\label{GEoS14}
\frac{B''\left(H_M\right)}{2} \left( \frac{d\delta N}{dt} \right)^2
\sim \beta_N m_N a_M^{m_N} H_M \left( t - t_M \right) \, ,
\end{equation}
which leads to
\begin{equation}
\label{GEoS15}
  - \frac{3\left( n - 2 \right)}{\kappa^2} \left( \frac{d\delta N}{dt} \right)^2
\sim m_N \left(  \frac{n \alpha \kappa^2}{6} \right)^{\frac{3}{2-n}}
\left( t - t_M \right) \, .
\end{equation}
Hence, we have that $\frac{d \delta N}{dt} \propto \sqrt{t_M - t}$,
and consequently as $H=H_M + \frac{d \delta N}{dt}$, the first derivative gives
\begin{equation}
\dot H \propto \left( t_M - t \right)^{- \frac{1}{2}}\ .
\end{equation}
And consequently a ``sudden'' singularity occurs at $t=t_M$.
\end{itemize}

Let us now analyze qualitatively the behavior for an analog model in early times,
when $a$ is small enough. As previously, we can express the first FLRW equation (\ref{GEoS1B})
together with (\ref{GEoS4}) and (\ref{GEoS6}) as
\begin{equation}
\label{GEoS16}
B(H) \sim \beta_1 a^{m_1}\, .
\end{equation}
Hence, depending on the sign of $m_1$ and the value of $n$, we can distinguish the following cases:
\begin{itemize}
\item For $m_1<0$, since $B(H)$ can be larger and larger as far as $n<2$, a Big Bang singularity
occurs at the beginning of the universe. On the other hand, if $n>2$, $B(H)$ has a maximum
as given in (\ref{GEoS10}) and consequently Eq.~(\ref{GEoS11}) provides a minimum $a_m$
for the scale factor,
\begin{equation}
\label{GEoS17}
a_m = \left( \frac{B\left( H_M \right)}{\beta_1} \right)^{\frac{1}{m_1}}
= \left( \frac{1}{\beta_1} \left(  \frac{n \alpha \kappa^2}{6} \right)^{\frac{2}{2-n}}
\right)^{\frac{1}{m_1}} \, ,
\end{equation}
which represents the counterpart of (\ref{GEoS12}). Then, through similar calculations,
we find $\dot H \propto \left( t - t_m \right)^{- \frac{1}{2}}$ near the time $t=t_m$,
where the scale factor $a$ takes a minimum value, $a=a_m$ and therefore there appears a
``sudden'' singularity at $t=t_m$. Note that the singularity is not a future one but
occurs at the beginning of the universe, i.e., the universe is generated by a Type II
singularity in this scenario.
\item For $m_1>0$, the corresponding perfect fluid is phantom-like, and the r.h.s. in (\ref{GEoS16})
decreases as the cosmological time goes far in the past and the initial state of the universe becomes
de Sitter, where $H$ is given by $H_0$ in (\ref{GEoS9}). In this scenario, there is no Big Bang and
the universe started at infinitely past.
\end{itemize}

\section{Early and late time universe for a more general class of equation of state \label{Sec4}}

We consider now a more general type of EoS than the ones described in (\ref{GEoS4}),
\begin{equation}
\label{GEoS18BBB}
f\left( H, a\right) = \alpha H^n + \beta H^l a^m \, .
\end{equation}
Here we assume $\alpha$ and $\beta$ are positive. Then, the FLRW equation (\ref{GEoS1}) takes the form,
\begin{equation}
\label{GEoS18}
C(H) \equiv \frac{3}{\kappa^2} H^{2-l} - \alpha H^{n-l}
= \beta a^m \, .
\end{equation}
Let us first analyze the properties of the function $C(H)$. Note that, as in the previous section, $C(H)$
vanishes at $H=H_0$, as given in (\ref{GEoS9}). Furthermore the properties of $C(H)$ depend on the values
of $l$ and $n$ and the relation among themselves.
\begin{itemize}
\item For $l<2$, the following cases might be raised:
\begin{itemize}
\item If $n>2$, there is a maximum at $H=H_{M_2}$, which is defined by
\begin{equation}
\label{GEoS19}
H_{M_2} \equiv \left( \frac{\left( n - l \right) \alpha \kappa^2}{3 \left( 2-l \right)}
\right)^\frac{1}{2-n} \, .
\end{equation}
\item For $n<2$ and $n-l>0$, there is a minimum and when $H$ grows,
$C(H)$ is positive and increases monotonically.
\item For $n<2$ and $n-l<0$, $C(H)$ is a monotonically increasing function and positive for $H>H_0$.
\end{itemize}
\item For $l>2$,
\begin{itemize}
\item For $n>2$, $C(H)$ is positive for $H<H_0$ and diverges as $H$ goes vanishes.
\item For $n<2$, $C(H)$ is positive when $H>H_0$ and has a maximum at $H= H_{M_2} > H_0$.
\end{itemize}
\end{itemize}
These properties for the function $C(H)$ suggest that the qualitative behavior for $l<2$ is similar to
the model analyzed in the previous section (\ref{GEoS4}). As above, we can analyze the universe expansion
in early and late times by analyzing the extreme values of the scale factor:

\

\noindent
\textbf{For $l<2$:}
\begin{itemize}
\item In late-time universe, when $a$ is large enough, we will have the following cases:
\begin{itemize}
\item If $m<0$, the universe goes to an asymptotically de Sitter spacetime as $H$ becomes constant,
$H=H_0$ (\ref{GEoS9}).
\item If $m>0$,
\begin{itemize}
\item In case $n>2$, a ``sudden'' singularity occurs.
\item In case $n<2$, there will be a Big Rip (Type I) singularity.
\end{itemize}
\end{itemize}
\item During the early universe, when $a$ is small enough, we will have
\begin{itemize}
\item For $m<0$,
\begin{itemize}
\item In case $n>2$, there is a ``sudden'' singularity as the initial singularity.
\item In case $n<2$, there is a Big Bang singularity.
\end{itemize}
\item For $m>0$, the universe tends to an asymptotically de Sitter spacetime where $H$ is a constant,
$H=H_0$ as given in (\ref{GEoS9}).
\end{itemize}
\end{itemize}

\noindent
\textbf{For $l>2$:}\\
On the other hand, for $l>2$, the model (\ref{GEoS18}) is a little bit different from
the model (\ref{GEoS4}) studied in the previous section.
\begin{itemize}
\item In late times, when $a$ is large enough, the following cases arise:
\begin{itemize}
\item For $m<0$, the universe goes to an asymptotically de Sitter spacetime with $H=H_0$,
as given in (\ref{GEoS9}).
\item For $m>0$,
\begin{itemize}
\item In case that $n>l>2$, the Hubble rate $H$ vanishes asymptotically and the space-time
becomes flat. Specifically, the Hubble rate (\ref{GEoS18}) can be approximated as
\begin{equation}
\label{GEoS18B}
H^2 \sim a ^{\frac{m}{2-l}}\, .
\end{equation}
Hence, the behavior of the universe is effectively given by a perfect fluid with an EoS parameter
given by $w=-1 - \frac{m}{3(2-l)}>-1$.
\item In case $n<2$, a sudden singularity occurs.
\end{itemize}
\end{itemize}
\item In early times, when $a$ is small enough, we will have
\begin{itemize}
\item For $m<0$,
\begin{itemize}
\item In case $n>l>2$, the Hubble rate $H$ vanishes as $a\rightarrow 0$ and the space-time
is nearly flat for small values of the scale factor, i.e., the initial state of the universe space-time is nearly Minkowski.
As in (\ref{GEoS18B}), the universe is effectively described by a perfect fluid with an EoS parameter given
by $w=-1 - \frac{m}{3(2-l)}>-1$, but there is  Big Bang singularity, since the Hubble parameter remains finite
but geodesics are not complete as $a\rightarrow 0$.
\item In case $n<2$, there will be a sudden singularity as the past singularity.
\end{itemize}
\item For $m>0$, the universe goes to asymptotically de Sitter spacetime where $H=H_0$ given in (\ref{GEoS9}).
\end{itemize}
\end{itemize}

We can go even beyond the model (\ref{GEoS18}) by considering an additional matter component in terms
of its dependence on the scale factor as follows,
\begin{equation}
B(H)=\frac{3}{\kappa^2}H^2-\alpha H^n=\beta_1 a^{m_1}+\beta_2 a^{m_2}\, .
\label{Model2a}
\end{equation}
By considering $m_1<0$ and $m_2>0$, the first term dominates in early epochs whereas
the second one becomes the dominant in late times. Hence, in early times, when $a$ is small, the first term dominates,
\begin{equation}
B(H)=\frac{3}{\kappa^2}H^2-\alpha H^n\sim\beta_1 a^{m_1}\, .
\label{Model2b}
\end{equation}
And we turn back to case analyzed in the previous section, such that for $n<2$, there is
an initial Big Bang singularity while for $n>2$, the scale factor is bounded by a minimum given in (\ref{GEoS17})
where a sudden singularity occurs.
In late times, for large values of the scale factor a similar behavior
to the previous section is also found, as the model (\ref{Model2a}) is approximated as,
\begin{equation}
B(H)=\frac{3}{\kappa^2}H^2-\alpha H^n\sim\beta_2 a^{m_2}\, .
\label{Model2c}
\end{equation}
Then, as far as $n<2$ a Big Rip singularity occurs in the future while $n>2$ leads to a future sudden singularity.
Hence, we might conclude that for the model (\ref{Model2a}), $n<2$ leads to a universe that starts
in a Big Bang singularity and ends in a Big Rip one, while $n>2$ provides an expansion starting and
ending through a sudden singularity.

In the next section, by following this procedure, we explore the construction of models that include
an early dark energy term and its behavior is analyzed.

\section{Early dark energy models \label{Sec5}}

Let us now reconstruct a model for early dark energy. By including all possible matter components
in the universe, the first FLRW equation is given by
\begin{equation}
\label{GEoS1BB}
\frac{3}{\kappa^2} H^2 = \rho_\mathrm{inf} + \rho_\mathrm{rad}
+ \rho_\mathrm{matter} + \rho_\mathrm{DE} + \rho_\mathrm{EDE} \, .
\end{equation}
Here $\rho_\mathrm{inf}$, $\rho_\mathrm{rad}$, $\rho_\mathrm{matter}$,
$\rho_\mathrm{DE}$, and $\rho_\mathrm{EDE}$ are the energy densities
corresponding to the inflaton, radiation, pressureless matter (baryonic and cold dark matter), dark energy,
and early dark energy, respectively.
We assume that the energy density for the EDE depends on the scale factor as
\begin{equation}
\label{EDE1}
\rho_\mathrm{EDE} = \frac{\rho_0 a^n}{a_0^{2n} + a^{2n}} \, .
\end{equation}
Here $\rho_0$ and $a_0$ are positive constants and we are assuming $n$ to be a positive integer.
Then, by inverting Eq.~(\ref{EDE1}), the scale factor can be expressed in terms of the energy density
for the EDE
\begin{equation}
\label{EDE1B}
a^n = - \frac{\rho_0}{2\rho_\mathrm{EDE}}
\pm \frac{1}{2} \sqrt{ \frac{\rho_0^2}{\rho_\mathrm{EDE}^2 } - 4 a_0^{2n} } \, .
\end{equation}
By using the continuity equation,
\begin{equation}
\label{EDE2}
0 = \dot \rho_\mathrm{EDE} + 3 H \left( \rho_\mathrm{EDE} + p_\mathrm{EDE}  \right) \, ,
\end{equation}
the following equation of state for the EDE is found
\begin{align}
\label{EDE3}
p_\mathrm{EDE} =& - \rho_\mathrm{EDE} + \frac{n}{3} \frac{\rho_0 a^n}{a_0^{2n} + a^{2n}}
 -  \frac{2n}{3} \frac{\rho_0 a^{3n}}{\left( a_0^{2n} + a^{2n}\right)^2 } \nn
=& \left( \frac{n}{3} - 1 \right) \rho_\mathrm{EDE}
 - \frac{n}{3} \rho_\mathrm{EDE}^2 \left( - \frac{\rho_0}{\rho_\mathrm{EDE}}
\pm \sqrt{ \frac{\rho_0^2}{\rho_\mathrm{EDE}^2 } - 4 a_0^{2n} } \right) \, .
\end{align}
In the early universe, when $a$ is small, $\rho_\mathrm{EDE}$ behaves as
$\rho_\mathrm{EDE} \propto a^n$, whereas in the late universe, when $a$ is large,
the EDE density can be approximated by  $\rho_\mathrm{EDE} \propto a^{-n}$.
Since $n$ is positive, $\rho_\mathrm{EDE}$ turns out negligible both in
the early universe as in the late universe if $n$ is chosen to be large enough.
Furthermore $\rho_\mathrm{EDE}$ has a positive maximum at $a=a_0$ and
therefore $\rho_\mathrm{EDE}$ behaves as an effective positive cosmological constant around
$a\sim a_0$ and consequently $\rho_\mathrm{EDE}$ plays the role of the so-called early dark energy.

As a second model for the EDE, similar to the previous one, we might consider the following EDE density,
\begin{equation}
\rho_\mathrm{EDE}=\rho_0 \e^{-(a-a_0)^{2n}}\, .
\label{EDEmodel2}
\end{equation}
Here we choose $n$ to be a positive integer.
In this case, the EDE density turns out negligible both for $a \ll a_0$ as for $a \gg a_0$,
such that as far as $a_0$ is fixed during the recombination epoch, the EDE model (\ref{EDEmodel2}) plays
its role just for $a\sim a_0$, when dominates, driving a short accelerating expansion.
For this case, the scale factor can be expressed in terms of the EDE density as
\begin{equation}
a-a_0=\left[\ln \left(\frac{\rho_0}{\rho_\mathrm{EDE}}\right)\right]^{1/2n}\, .
\end{equation}
And the corresponding EoS can be obtained through the continuity equation (\ref{EDE2}), leading to:
\begin{equation}
p_\mathrm{EDE}=-\rho_\mathrm{EDE}+\frac{2}{3}n\rho_\mathrm{EDE}
\left[\ln \left(\frac{\rho_0}{\rho_\mathrm{EDE}}\right)\right]^{(2n-1)/2n}
\left\{a_0+\left[\ln \left(\frac{\rho_0}{\rho_\mathrm{EDE}}\right)\right]^{1/2n}\right\}\, .
\end{equation}

As a third model for the EDE and inspired in the ones considered in the previous sections,
we might express the EDE density as a function of the Hubble parameter,
\begin{equation}
\label{EDE4}
\rho_\mathrm{EDE} = \frac{\rho_0 H^n}{H_0^{2n} + H^{2n}} \, .
\end{equation}
Here $\rho_0$ is a positive constant and $n$ is a positive integer, again.
By assuming that $H \gg H_0$ in the early universe, the EDE density (\ref{EDE4}) can be approximated as
$\rho_\mathrm{EDE} \propto H^{-n}$ and consequently becomes negligible in early times.
Similarly, in the late universe, when $H \ll H_0$, we find  $\rho_\mathrm{EDE} \propto H^n$,
such that becomes also negligible in the late universe. Moreover, $\rho_\mathrm{EDE}$ has
a positive maximum at $H=H_0$ and consequently $\rho_\mathrm{EDE}$ behaves as
a positive cosmological constant over the period when $H\sim H_0$, playing the the role of early dark energy.
By the continuity equation (\ref{EDE2}), the EoS for the EDE model (\ref{EDE4}) yields
\begin{equation}
\label{EDE5}
p_\mathrm{EDE} = - \rho_\mathrm{EDE} + \frac{n\dot H}{3H^2} \frac{\rho_0 H^n}{H_0^{2n} + H^{2n}}
 -  \frac{2n \dot H}{3H^2} \frac{\rho_0 H^{3n}}{\left( H_0^{2n} + H^{2n}\right)^2 } \, .
\end{equation}
Then, by using (\ref{EDE4}), we can rewrite (\ref{EDE5}) as follows,
\begin{equation}
\label{EDE6}
p_\mathrm{EDE} = \left( - 1 + \frac{n\dot H}{3H^2}\right) \rho_\mathrm{EDE}
 -  \frac{2n H^{n-2} \dot H}{3\rho_0} \rho_\mathrm{EDE} ^2 \, ,
\end{equation}
which provides the EoS for the EDE model (\ref{EDE4}) in terms of $H$ and $\dot H$.

Finally, let us analyze the models that we will fit below to observational data.
As a more general EoS that unifies dark energy and early dark energy,
we consider here the following model
\begin{equation}
\label{O1}
p = - \rho + \gamma_1 \rho^{l_1} + \gamma_2 \rho^{l_2} + \alpha_0 H^{n_0} + \beta a^m \, .
\end{equation}
Here $\gamma_1$, $\gamma_2$, $\alpha_0$, and $\beta$ are constants.
Then, by using the FLRW equations $H^2=\frac{\kappa^2}{3}\rho$ and
$\dot H = - \frac{\kappa^2}{2} \left( p + \rho \right)$, the EoS (\ref{O1})
can be expressed in terms of the Hubble parameter as follows,
\begin{equation}
\label{O2}
 - \frac{2}{\kappa^2} \dot H = \alpha_1 H^{n_1} + \alpha_2 H^{n_2} + \alpha_0 H^{n_0} + \beta a^m \, ,
\end{equation}
where we have renamed the corresponding parameters as
\begin{equation}
\label{O3}
n_1 \equiv 2l_1 \, , \quad n_2 \equiv 2l_2 \, , \quad
\alpha_1 \equiv \gamma_1 \left( \frac{\kappa^2}{3} \right)^{n_1} \, , \quad
\alpha_2 \equiv \gamma_2 \left( \frac{\kappa^2}{3} \right)^{n_2} \, .
\end{equation}
When considering asymptotic behaviors, as in the late or in the early universe, one power
of the Hubble parameter dominates over the rest, which we can rename as
$\alpha H^n$. Then, Eq.~(\ref{O2}) can be approximated as
\begin{equation}
\label{O4}
 - \frac{2}{\kappa^2} \dot H = \alpha H^{n} + \beta a^m \, .
\end{equation}
Here there still appear three terms but in the late or early universe, one term might be
negligible and two terms balance with each other. Let us analyze each case:
\begin{itemize}
\item In case that the term $- \frac{2}{\kappa^2} \dot H$ can be neglected,
Eq.~(\ref{O4}) is approximated as
\begin{equation}
\label{O5}
H^2 \sim \left( - \frac{\beta}{\alpha} \right)^{\frac{2}{n}} a^\frac{2m}{n} \, .
\end{equation}
The universe expands as if there were a perfect fluid with a constant EoS parameter
$w=-1 - \frac{2m}{3n}$.
\item  In case that the term $\beta a^m$ is neglected in comparison to the others,
Eq.~(\ref{O4}) can be approximated as
\begin{equation}
\label{O6}
 - \frac{2}{\kappa^2} \dot H \sim \alpha H^{n} \, ,
\end{equation}
which after integrating leads to
\begin{equation}
\label{O7}
H \sim \left\{ \frac{ \alpha \kappa^2(n-1)}{2} \left( t - t_0
\right)\right\}^{-\frac{1}{n-1}}\, .
\end{equation}
Here $t_0$ is a constant of the integration. If $n>1$,
 a Big Rip singularity occurs in the late universe or a Big Bang singularity in case of the early universe.
\item In case that the term $\alpha H^{n}$ can be neglected, Eq.~(\ref{O4}) can be approximated as
\begin{equation}
\label{O8}
 - \frac{2}{\kappa^2} \dot H \sim \beta a^m \, .
\end{equation}
By using the number of e-foldings $a$ as $a=\e^N$, Eq.~(\ref{O8}) can be rewritten as
\begin{equation}
\label{O9}
 - \frac{2}{\kappa^2} \ddot N \sim \beta \e^{mN} \, .
\end{equation}
Then, by multiplying the equation by $\dot N=H$, and after integrating, we obtain
\begin{equation}
\label{O10}
 - \frac{1}{\kappa^2} {\dot N}^2 \sim \frac{\beta}{m} \e^{mN} + C\, ,
\end{equation}
with $C$ being a constant of integration. And finally Eq.~(\ref{O10}) can be expressed as
\begin{equation}
\label{O11}
\frac{3}{\kappa^2} H^2 \sim - \frac{3\beta}{m} a^m - 3 C\, .
\end{equation}
The universe behaves as in the presence of a perfect fluid with a constant EoS parameter
$w=-1 - \frac{2m}{3}$ and cosmological constant $-3C$.
\end{itemize}


As a second model for the EoS that we will use for comparing to observational data in the section below,
we might consider an EoS  that contains both
powers of the scale factor and of the Hubble parameter
\begin{equation}
\label{EoS1} p = - \rho + \gamma_1 \rho^{l_1} +  b\, a^m  H^{n_0} \, .
\end{equation}
Here $\gamma_1$ and $b$ are constants.
This model belongs to the more general class given in (\ref{GEoS1}).
By using the FLRW equations, the EoS can be expressed as
\begin{equation}
\label{E2}
 - \frac{2}{\kappa^2} \dot H = \alpha_1 H^{n_1} + b\, a^m  H^{n_0}\ ,
\end{equation}
which can also be analyzed within 3 asymptotic approximations
\begin{itemize}
\item In case that the term $- \frac{2}{\kappa^2} \dot H$ can be neglected,
Eq.~(\ref{E2}) is approximated as
\begin{equation}
H^{n_1-n_0} \sim  - \frac{b}{\alpha_1} a^m\ .
\end{equation}
This describes an expanding universe with a perfect fluid with a constant EoS parameter
$w=-1 - \frac{2}{3}\frac{m}{n_1-n_0}$.
\item  In case that the term $ b\, a^m  H^{n_0}$ is neglected, Eq.~(\ref{E2}) is reduced
to the form (\ref{O6}), that can be integrated similarly to Eq.~(\ref{O7}), leading to
\[
H \simeq \left\{ \frac{ \alpha_1 \kappa^2(n_1-1)}{2} \left( t - t_0
\right)\right\}^{-\frac{1}{n_1-1}}\, .
\]
If $n_1>1$, this solution also contain a Big Rip singularity for the late universe or
a Big Bang singularity for the early universe.
\item In case that the term $\alpha_1 H^{n_1}$ can be neglected, Eq.~(\ref{E2}) is reduced to
\[
H^{1-n_0} \dot H \simeq- \frac12 b {\kappa^2} a^{m-1}\dot a  \, .
\]
which after integrating, yields
\begin{equation}
\label{E3} H^{2-n_0} \simeq  - \frac{2-n_0}{2m}b{\kappa^2} a^m + C\, .
\end{equation}
As can be easily noted, one of the asymptotic behaviors of the previous model given
in Eq.~(\ref{O11}) is a particular case of Eq.~(\ref{E3}) for $n_0=0$. Nevertheless. the
expression (\ref{E3}) can not be interpreted in general as Eq.~(\ref{O11}),
but the corresponding value for the parameters has to be provided.
\end{itemize}

In the next section, we compare these last two models for (early) dark energy
with several sources of observational data.

\section{Observational tests} \label{Sec6}

Viability of the above models is now analyzed by comparing their predicting power
with observational data, including estimations of the Hubble parameter $H(z)$,
Supernovae Type Ia (SNe Ia), baryon acoustic
oscillations (BAO) and cosmic microwave background radiation (CMB) distances.
For this purpose we will use some techniques developed in some previous papers
\cite{Odintsov:2017qif,Odintsov:2018qug,Odintsov:2020voa,Sharov:2015ifa,Pan:2016ngu}
and concentrate on the models given in the previous section by (\ref{O1}) and  (\ref{EoS1}).\\

Here we also include the matter components corresponding to radiation and pressureless matter,
which are denoted by: $\rho_r$ for radiation,
$\rho_m\equiv\rho_\mathrm{matter}$ for baryons with dark matter (together) and
$\rho_\mathrm{DE}$ for (early) dark energy, with $\rho\equiv\rho_\mathrm{DE}$, $p\equiv p_\mathrm{DE}$
as given in (\ref{O1}) or (\ref{EoS1}), such that the FLRW equations yield:
\begin{equation}
\label{OT1}
\frac{3}{\kappa^2} H^2 = \rho_r + \rho_m +\rho\ ,\quad
\rho\equiv\rho_\mathrm{DE} \, .
\end{equation}

We  assume that any component does not interact with each other
and satisfy the continuity equation independently,
\begin{equation}
\dot{\rho}_i+ 3 H (p_i+\rho_i)= 0\,,
\label{cont}
\end{equation}
which for pressureless matter and radiation leads to
\begin{equation}
\rho_m=\rho_m^0a^{-3} \, ,\quad \rho_r=\rho_r^0a^{-4}\, .
\label{rhomr}
\end{equation}
Here the subindex ``$^0$'' refers to magnitudes measured at the present time $t_0$, in particular,
the Hubble constant is given by $H_0=H(t_0)$,  while for the scale factor we assume $a(t_0)=1$.
The evolution for dark energy density  can be obtained by solving the continuity equation (\ref{cont})
that can be rewritten as
\begin{equation}
\frac{d\rho}{d \ln a} = -3(p+\rho)\, ,
\label{cont2}
\end{equation}
which can be integrated numerically (in general) for a particular EoS as the ones analyzed in
the previous sections. For the model (\ref{EoS1}), the continuity equation (\ref{cont2})
may be rewritten as
\begin{equation}
\frac{d\Omega_\mathrm{DE}}{d \ln a} 
= -3\left[A\Omega_\mathrm{DE}^{\,l_1} + Ba^m\left(\frac{H}{H_0} \right)^{2\beta}\right]\,.
\label{Eqcont}
\end{equation}
Here  $\Omega_\mathrm{DE}$ is defined as usual as the ratio of dark energy density and
critical density, in the same way that applies to the rest of the components
\begin{equation}
\Omega_\mathrm{DE}=\frac{\rho_\mathrm{DE}}{\rho_\mathrm{cr}}
=\frac{\kappa^2\rho_\mathrm{DE}}{3H_0^2} \, ,\quad
\Omega_i=\frac{\rho_i}{\rho_\mathrm{cr}}\, ,
\end{equation}
where $A$, $B$, and $\beta$ are the parameters that describe the model (\ref{EoS1})
but redefined in such a way to keep them dimensionless
\begin{equation}
A=\gamma_1\rho_\mathrm{cr}^{l_1-1}\ , \quad B=bH_0^{n_0}/\rho_\mathrm{cr}\ , \quad n_0=2\beta\, .
\end{equation}
Then, we can integrate numerically the equation (\ref{Eqcont}) together with the equations (\ref{OT1}) and
(\ref{rhomr}) from $a=1$ ($t=t_0$) with the initial conditions
\begin{equation}
\Omega_\mathrm{DE}^0=1- \Omega_m^0- \Omega_r^0\, ,
\label{ini}
\end{equation}
which is the constraint equation (\ref{OT1}) evaluated at $t=t_0$ with $\Omega_i^0=\Omega_i(t_0)$.
This approach can be applied to any EoS for dark energy. In particular, for the model (\ref{O1})
in a simpler version by assuming $\gamma_2=0$, the continuity equation yields
\begin{equation}
\frac{d\Omega_\mathrm{DE}}{d \ln a} 
= -3\left[ A\Omega_\mathrm{DE}^{\,l_1} + B\left( \frac{H}{H_0} \right)^{2\beta}+ Ca^m\right]\,.
\label{Eqcont1}
\end{equation}
Note that both models (\ref{Eqcont}) and (\ref{Eqcont1}) contain a large number $N_p$ of
free parameters to be fit with observational data. Large $N_p$  is a serious drawback
for any model in comparison with other cosmological scenarios, as the $\Lambda$CDM model, from the point
of view of information criteria \cite{Akaike74,Schwarz78}. In order to reduce the number
of the free parameters, we are not considering $\Omega_r^0$ as an independent parameter
but the ratio among cold matter and radiation is fixed as provided by Planck
\cite{Ade:2013zuv,Odintsov:2018qug,Odintsov:2020voa}
\begin{equation}
X_r=\frac{\rho_r^0}{\rho_m^0}=\frac{\Omega_r^0}{\Omega_m^0}=2.9656\times 10^{-4}\,.
\label{Xrm}
\end{equation}
This value is rather small, so while is essential for CMB observational data for redshifts $z\simeq1000$,
becomes negligible for SNe Ia, $H(z)$ and BAO observations in the range $0<z\le2.36$.
In order to simplify the models even more, we also fix the value $l_1=1$ for both models,
as our calculations and fittings show that both models (\ref{Eqcont}) and (\ref{Eqcont1}) depend weakly on
$l_1$, similarly as was shown in Ref.~\cite{Odintsov:2020voa} for an analog model. Note also that for case
$l_1=1$, the models (\ref{Eqcont}) and (\ref{Eqcont1}) can be reduced to the $w$CDM model with
the EoS parameter $w=A-1$ as far as one fixes $B=0$ (and $C=0$).

Hence, by fixing the ratio (\ref{Xrm}) and the value $l_1=1$, we have  the following set
of free parameters for the two models
\begin{equation}
\begin{array}{ll}
\Omega_m^0, \; A,\; B,\;m,\;\beta,\; H_0, & \mbox{ \  model (\ref{Eqcont})}\,,\\
\Omega_m^0, \; A,\; B,\;\beta,\; C,\;m,\; H_0, \;\;& \mbox{ \ model (\ref{Eqcont1})}\,.
\rule{0mm}{1.2em}\end{array}
\label{6param}
\end{equation}

Note that the model (\ref{Eqcont}) has  $N_p=6$ free parameters, whereas the
model (\ref{Eqcont1}) owns $N_p=7$. Both significantly exceed $N_p=3$ for the
$w$CDM model ($\Omega_m^0$, $H_0$ and $w$) and $N_p=2$
for the flat $\Lambda$CDM model ($\Omega_m^0$ and $H_0$).
However, below we
reduce the effective number $N_p$ by considering the Hubble constant $H_0$ as a nuisance parameter.

\subsection{Type Ia supernovae data}

For testing our models we use several observational datasets, including the latest
Pantheon sample \cite{Scolnic:2017caz} of Supernovae Ia (SNe Ia), estimations of the
Hubble parameter $H(z)$, observational data from baryon acoustic oscillations
(BAO) and cosmic microwave background radiation (CMB)
\cite{Ade:2013zuv,Ade:2015xua,Aghanim:2018eyx}.

For SNe Ia we use the largest most recent catalogue, the so-called Pantheon sample
\cite{Scolnic:2017caz} including $n_{\mbox{\scriptsize SN}}=1048$ SNe Ia data points with
redshifts $0< z_i\le2.26$ and distance moduli $\mu_i^\mathrm{obs}$. For every model we calculate
the theoretical value of the luminosity distances $D_L(z; \lambda_1,\lambda_2,\dots)$
and distance modulus $\mu_i^\mathrm{th}$ for each set of the free parameters $\lambda_k$ given
in (\ref{6param}) for each model
\begin{equation}
\mu^\mathrm{th} \left( z;\lambda_k \right)
= 5 \log_{10} \frac{D_L \left( z;\lambda_k \right)}{10\, \mbox{pc}}\, ,
\quad  D_L \left( z; \lambda_k \right)
= c (1+z) \int_0^z \frac{d\tilde z}{H \left( \tilde z; \lambda_k \right)}\, .
\label{muD}
\end{equation}
Then, the corresponding $\chi^2$ function is obtained
\begin{equation}
\chi^2_{\mathrm{SN}} \left( \Omega_m^0,A,\dots \right)
=\min\limits_{H_0} \sum_{i,j=1}^{1048}
\Delta\mu_i\left(C_{\mathrm{SN}}^{-1}\right)_{ij} \Delta\mu_j\, ,\quad
\Delta\mu_i=\mu^\mathrm{th} \left(z_i;\lambda_k \right)-\mu^\mathrm{obs}_i\, .
\label{chiSN}
\end{equation}
Here  $C_{\mathrm{SN}}$ is the $1048\times1048$ covariance matrix \cite{Scolnic:2017caz}.
For each set of the model parameters (\ref{6param}), we solve the system of
equations formed by (\ref{OT1}), (\ref{rhomr}), and the corresponding continuity equation
for each model given in (\ref{Eqcont}) and (\ref{Eqcont1}), obtaining the
Hubble parameter $H(z)=H \left( z,\lambda_k \right)$ and the corresponding luminosity distance
and distance modulus (\ref{muD}). Finally, we
marginalize  $\chi^2_{\mathrm{SN}}$ over the nuisance parameter $H_0$
\cite{Odintsov:2017qif,Odintsov:2018qug,Odintsov:2020voa,Sharov:2015ifa,Pan:2016ngu}.

\subsection{BAO data}

From the Baryon Acoustic Oscillations (BAO) data, provided by the analysis of galaxy
clustering, we can extract two magnitudes \cite{Eisenstein:2005su}
\begin{equation}
d_z(z)= \frac{r_s(z_d)}{D_V(z)}\, ,\quad
A(z) = \frac{H_0\sqrt{\Omega_m^0}}{cz}D_V(z)\, ,
\label{dzAz}
\end{equation}
and then compare with the corresponding theoretical predictions, where
\[
D_V(z)=\left[\frac{cz D_M^2(z)}{H(z)} \right]^{1/3}\, ,\quad
D_M(z)=\frac{D_L(z)}{1+z}= c \int_0^z \frac{d\tilde z}{H (\tilde z)}\, .
\]

In Eq.~(\ref{dzAz}), $r_s(z_d)$ is the comoving sound horizon at the end of the baryon
drag era $z_d$, this value can essentially vary in models with EDE or modified EoS
\cite{Aylor:2019}, such that we calculate it numerically through the following integral:
  \begin{equation}
 r_s(z)=  \int_z^{\infty}
\frac{c_s(\tilde z)}{H (\tilde z)}\,d\tilde z=\frac1{\sqrt{3}}\int_0^{1/(1+z)}\frac{da}
 {a^2H(a)\sqrt{1+\big[3\Omega_b^0/(4\Omega_\gamma^0)\big]a}}\ ,
  \label{rs2}\end{equation}
Here we have used Eq.~(\ref{Xrm}), $\rho_\nu=N_\mathrm{eff}(7/8)(4/11)^{4/3}\rho_\gamma$
 with $N_\mathrm{eff} = 3.046$ and other estimations from Planck collaboration 2018 data \cite{Aghanim:2018eyx}.

Moreover, we use 17 BAO data points for $d_z(z)$ and 7 data points for $A(z)$ from
Refs.~\cite{Percival:2009xn,Beutler:2011hx,Blake:2011en,Padmanabhan:2012hf,Chuang:2012qt,
Chuang:2013hya,Ross:2014qpa,Anderson:2013zyy,Oka:2013cba,Font-Ribera:2013wce,
Delubac:2014aqe} as in other previous papers
\cite{Odintsov:2017qif,Odintsov:2018qug,Odintsov:2020voa}, where these data points are
tabulated. The $\chi^2$ function for the BAO data (\ref{dzAz}) has the form
\begin{equation}
\chi^2_{\mathrm{BAO}}(\Omega_m^0,A,\dots)=\Delta d\cdot C_d^{-1}(\Delta d)^T
+ \Delta { A}\cdot C_A^{-1}(\Delta { A})^T\, ,
\label{chiB}
\end{equation}
where $\Delta d$, $\Delta A$ are vectors given by
\begin{equation}
\Delta d_i=d_z^\mathrm{obs}(z_i)-d_z^\mathrm{th}(z_i,\dots)\, , \quad
\Delta A_i=A^\mathrm{obs}(z_i)-A^\mathrm{th}(z_i,\dots)\, ,
\end{equation}
while $C_{d}$ and $C_{A}$ are the covariance matrices for correlated BAO data
\cite{Sharov:2015ifa,Percival:2009xn,Beutler:2011hx,Blake:2011en,Padmanabhan:2012hf,
Chuang:2012qt,Chuang:2013hya,Ross:2014qpa,Anderson:2013zyy,Oka:2013cba,
Font-Ribera:2013wce,Delubac:2014aqe}.

\subsection{$H(z)$ data}

Here we use the Hubble parameter data $H(z)$ estimated by the method of differential
ages $\Delta t$ for galaxies with small differences $\Delta z$ in redshifts (cosmic
chronometers), where the values $H(z)$ can be extracted through the relation
\[ 
H (z)= \frac{\dot{a}}{a} \simeq -\frac{1}{1+z}
\frac{\Delta z}{\Delta t}\,.
\]
We use $N_H=31$ data points $H^\mathrm{obs}(z_i)$ of cosmic chronometers from
Refs.~\cite{Simon:2004tf,Stern:2009ep,Moresco:2012jh,Zhang:2012mp,Moresco:2015cya,
Moresco:2016mzx,Ratsimbazafy:2017vga} in the redshift interval $0<z<2$.
These measurements are not
correlated with the BAO data points \cite{Percival:2009xn,Beutler:2011hx,Blake:2011en,
Padmanabhan:2012hf,Chuang:2012qt,Chuang:2013hya,Ross:2014qpa,Anderson:2013zyy,
Oka:2013cba,Font-Ribera:2013wce,Delubac:2014aqe}.
The $\chi^2$ function for $H(z)$ data is
\begin{equation}
\chi^2_{H}=\min\limits_{H_0} \sum_{i=1}^{N_H} \left[\frac{H^\mathrm{obs}(z_i)
 -H^\mathrm{th}(z_i; \lambda_k)}{\sigma_{H,i}}\right]^2 \, .
\label{chiH}
\end{equation}

\subsection{CMB data}

Unlike the above data, the CMB observations are related to the photon-decoupling epoch
at $z_*\simeq1090$ ($z_*=1089.80 \pm0.21$ \cite{Aghanim:2018eyx}),
  where the radiation density $\rho_r(z)$ is essential. We use the
following CMB observational parameters
\[ 
\mathbf{x}=\left(R,\ell_A,\omega_b \right)\, ,\quad
R=\sqrt{\Omega_m^0}\frac{H_0D_M(z_*)}c\, ,\quad
\ell_A=\frac{\pi D_M(z_*)}{r_s(z_*)}\, ,
\quad\omega_b=\Omega_b^0h^2
\] 
with the estimations \cite{Chen:2018dbv}
\begin{equation}
\mathbf{x}^\mathrm{Pl}=\left( R^\mathrm{Pl},\ell_A^\mathrm{Pl},\omega_b^\mathrm{Pl} \right)
=\left( 1.7428\pm0.0053,\;301.406\pm0.090,\;0.02259\pm0.00017 \right) \, ,
\label{CMBpriors}
\end{equation}
They are extracted from Planck collaboration 2018 data \cite{Aghanim:2018eyx} with free
amplitude for the lensing power spectrum.

The expression $r_s(z_*)$ is calculated by the integral (\ref{rs2}), where  for the
value $z_*$ we use the fitting formula given in Refs.~\cite{Chen:2018dbv,HuSugiyama95}.
 The
current baryon fraction $\Omega_b^0$ is considered as the nuisance parameter to
marginalize over (together with $H_0$). The corresponding $\chi^2$ function is
\begin{equation}
\chi^2_{\mathrm{CMB}}=\min_{\omega_b,H_0}\Delta\mathbf{x}\cdot
C_{\mathrm{CMB}}^{-1}\left( \Delta\mathbf{x} \right)^{T}\, ,\quad
\Delta \mathbf{x}=\mathbf{x}-\mathbf{x}^\mathrm{Pl}\,.
\label{chiCMB}
\end{equation}
The covariance matrix $C_{\mathrm{CMB}}=\left| \tilde C_{ij}\sigma_i\sigma_j \right|$
and other details are described in Refs.~\cite{Odintsov:2020voa} and
\cite{Chen:2018dbv}.

\section{Results and discussion}
\label{Sec7}

\begin{figure}[th]
\centerline{ \includegraphics[scale=0.66,trim=5mm 0mm 2mm -1mm]{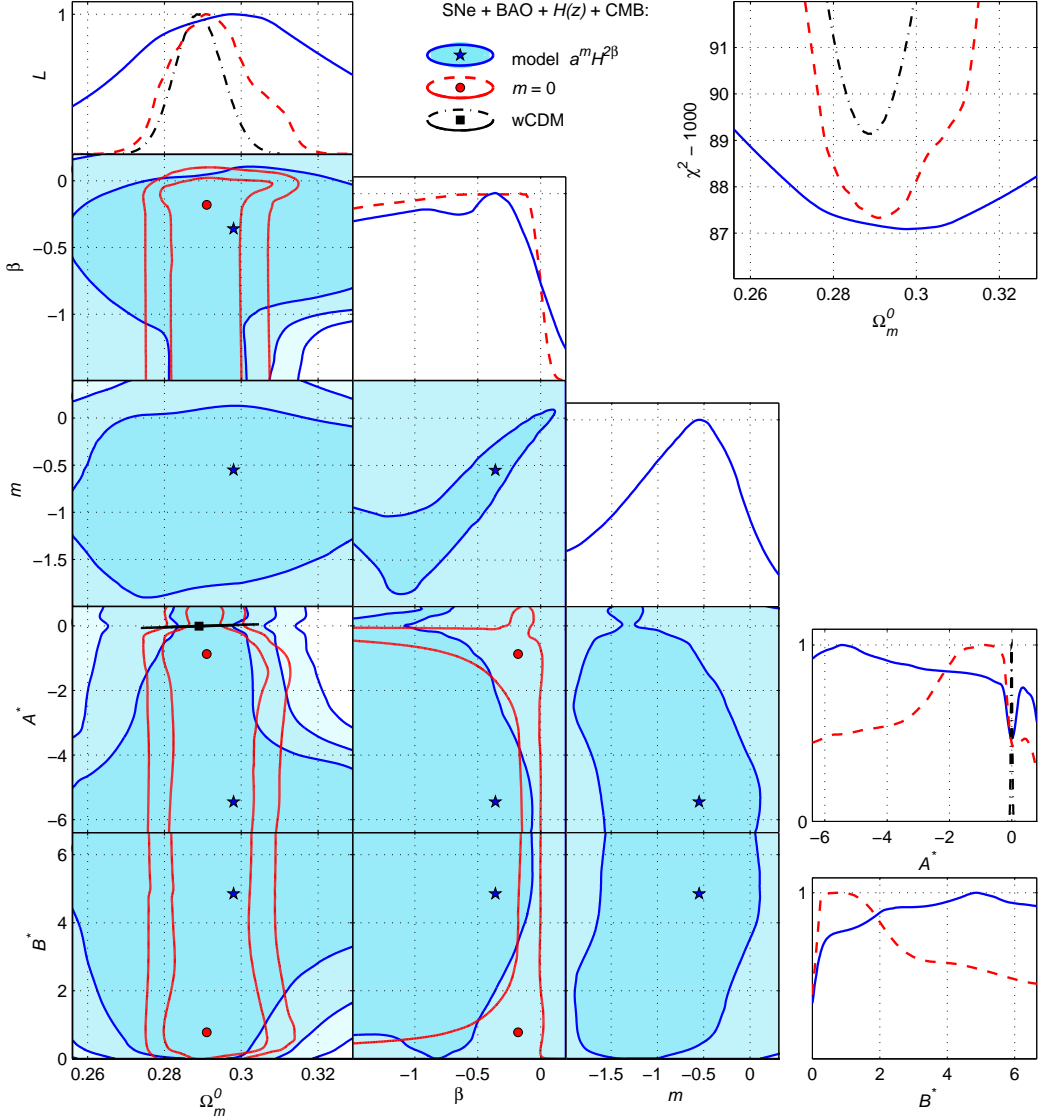}}
\caption{Contours plots, likelihoods and  $\chi^2_{\mathrm{tot}}(\Omega_m^0)$ for the
model (\ref{Eqcont}) (blue). The particular cases $m=0$ (red) and  $w$CDM model (black)
are also depicted.}
  \label{F1}
\end{figure}

Let us now analyze and fit the models (\ref{Eqcont}) and (\ref{Eqcont1}) to the above SNe Ia, BAO,
$H(z)$, and CMB datasets and obtain the corresponding constraints on the free model parameters.
The total $\chi^2$ function is obtained by the sum of the partial ones
\cite{Odintsov:2018qug,Odintsov:2020voa}:
\begin{equation}
\chi^2_{\mathrm{tot}}=\chi^2_{\mathrm{SN}}+\chi^2_H+\chi^2_{\mathrm{BAO}}+\chi^2_{\mathrm{CMB}}\, ,
\label{chitot}
\end{equation}
which contain the fittings to all the observational datasets and will provide the corresponding
confidence regions in the parameter spaces (\ref{6param}).
The best fits for the free model parameters $\lambda_k$ are obtained by using the
one-parameter distributions $\chi^2_{\mathrm{tot}}(\lambda_k)$ and the corresponding
likelihoods:
\begin{equation}
\mathcal{L}(\lambda_k)\propto \exp\left[-\frac12\chi^2_{\mathrm{tot}}(\lambda_k)\right]\, ,
\label{likelihood}
\end{equation}
where we assume a Gaussian distribution for the free parameters.  The results for the model (\ref{Eqcont})
are shown in Fig.~\ref{F1}, where the contour plots
are depicted together with the likelihoods for each parameter
separately. The blue filled contour plots denote $1\sigma$ (68.27\%), $2\sigma$
(95.45\%) and $3\sigma$ (99.73\%) confidence levels (CL) for two-parameter distributions, where in each panel
we minimize $\chi^2_{\mathrm{tot}}$ over all the other parameters. For
example, in $ \Omega_m^0-\beta$ plane the contours are drawn for
\[
 \chi^2_{\mathrm{tot}}(\Omega_m^0, \beta)=\min\limits_{A,B,m,H_0} \chi^2_{\mathrm{tot}}(\Omega_m^0,A,B,m,\beta,H_0)\,.
\]
Note that $H_0$ and $\omega_b$ are considered as nuisance parameters, so we minimise the
likelihood over them in the corresponding fittings. The red contour plots depict
$1\sigma$  CL (and $2\sigma$ CL in the left panels) for the particular case $m=0$ of
this model, which was also considered in Ref.~\cite{Odintsov:2020voa}. Recall that the
model (\ref{Eqcont}) under the condition $B=0$ turns out $w$CDM model with $w=A-1$, such
that the results for this case are also shown in  Fig.~\ref{F1}. The corresponding
minimums for $\chi^2_{\mathrm{tot}}$ in these cases are labeled as stars and circles.

For convenience, in the bottom panels of Fig.~\ref{F1} and in Table~\ref{Estim} we have
redefined the free parameters $A$ and $B$ as
\begin{equation}
A=\sinh A^*\, , \quad B=\sinh B^*\,, \label{ABsh}
\end{equation}

For the model  (\ref{Eqcont}) and its particular cases $m=0$ and $B=0$ in the top-right
panel, one-parameter distributions $\chi^2_{\mathrm{tot}}(\Omega_m^0)$ are depicted by minimizing over
all the other parameters. Here we can compare the absolute minimum of
$\chi^2_{\mathrm{tot}}$ for these cases, tabulated below in Table~\ref{Estim} with the best fits
for the model free parameters.
One can see that the model with $m=0$ (red lines) is rather successful from this point
of view in comparison with the  $w$CDM model ($B=0$). However, if we compare with the
case $m=0$,  the model (\ref{Eqcont}) in its general case (blue lines) just provides an slightly smaller $\min\chi^2_{\mathrm{tot}}$ but a larger number of free
parameters $N_p=6$ and larger errors. Actually, the better fits in terms of the $chi^2$ might be connected with the large best fits values for $|A^*|$ and $B^*$ in
the general case of the model (\ref{Eqcont}). For the case $m=0$, the best fits remain smaller, $A^*\simeq-0.87$, $B^*\simeq0.78$ . These specific features lead to different $1\sigma$
 ranges in $\Omega_m^0$, as shown in  Table~\ref{Estim}. \\

\begin{table}[bh]
 \centering
 {\begin{tabular}{||l|c|c|c|c|c|c|l||}  \hline
 Model  & $\Omega_m^0$& $A^*$ & $B^*$ & $m$ &$\beta$ & $C$&$ \min\chi^2_{\mathrm{tot}}\,/\,$d.o.f  \\ \hline\hline
Eq.\,(\ref{Eqcont})& $0.298_{-0.029}^{+0.029}$ & $-5.45_{-\infty}^{+5.30}$  & 
$4.85_{-4.66}^{+\infty}$  & $-0.55_{-0.74}^{+0.45}$ & $-0.36_{-1.35}^{+0.32}$& -& 1087.09 / 1101\rule{0pt}{1.2em}  \\
 \hline
 Eq.\,(\ref{Eqcont1})& $0.283_{-0.013}^{+0.029}$ &  $-2.48_{-3.05}^{+3.22}$  & $2.27_{-1.38}^{+1.88}$
 & $19.7_{-18.4}^{+38.5}$ & $-0.114_{-0.133}^{+0.085}$  & $-0.86_{-5.44}^{+0.98}$& 1086.36 / 1100\rule{0pt}{1.2em}  \\
 \hline
$C=0$ &  $0.291_{-0.010}^{+0.011}$ & $-0.87_{-2.29}^{+0.76}$  &
$0.78_{-0.66}^{+2.04}$ & 0&  $-0.18_{-1.85}^{+0.17}$ & 0& 1087.34 /1102\rule{0pt}{1.2em}  \\
 \hline
$w$CDM &  $0.289_{-0.006}^{+0.006}$ &$-0.007_{-0.023}^{+0.024}$ & - & - & - & -& 1089.14 / 1104 \rule{0pt}{1.2em} \\
  \hline \end{tabular}
\caption{Best fits from SNe Ia, $H(z)$, BAO and CMB data for the models (\ref{Eqcont}),
(\ref{Eqcont1}) and their particular cases $m=C=0$  and  $w$CDM ($B=C=0$).}
 \label{Estim}}
\end{table}

Remind that under the condition $B=0$ the models (\ref{Eqcont}) is reduced to the $w$CDM
model with $w=A-1$. In this case the best fitted parameter yields:
 \be A\simeq
A^*=-0.007_{-0.023}^{+0.024}\quad \rightarrow\quad w=-1.007_{-0.023}^{+0.024}\ . \ee
And the EoS lies  mostly in the phantom regime, although includes quintessential values within its 1-$\sigma$ domain.\\

\begin{figure}[th]
\centerline{ \includegraphics[scale=0.66,trim=5mm 0mm 2mm -1mm]{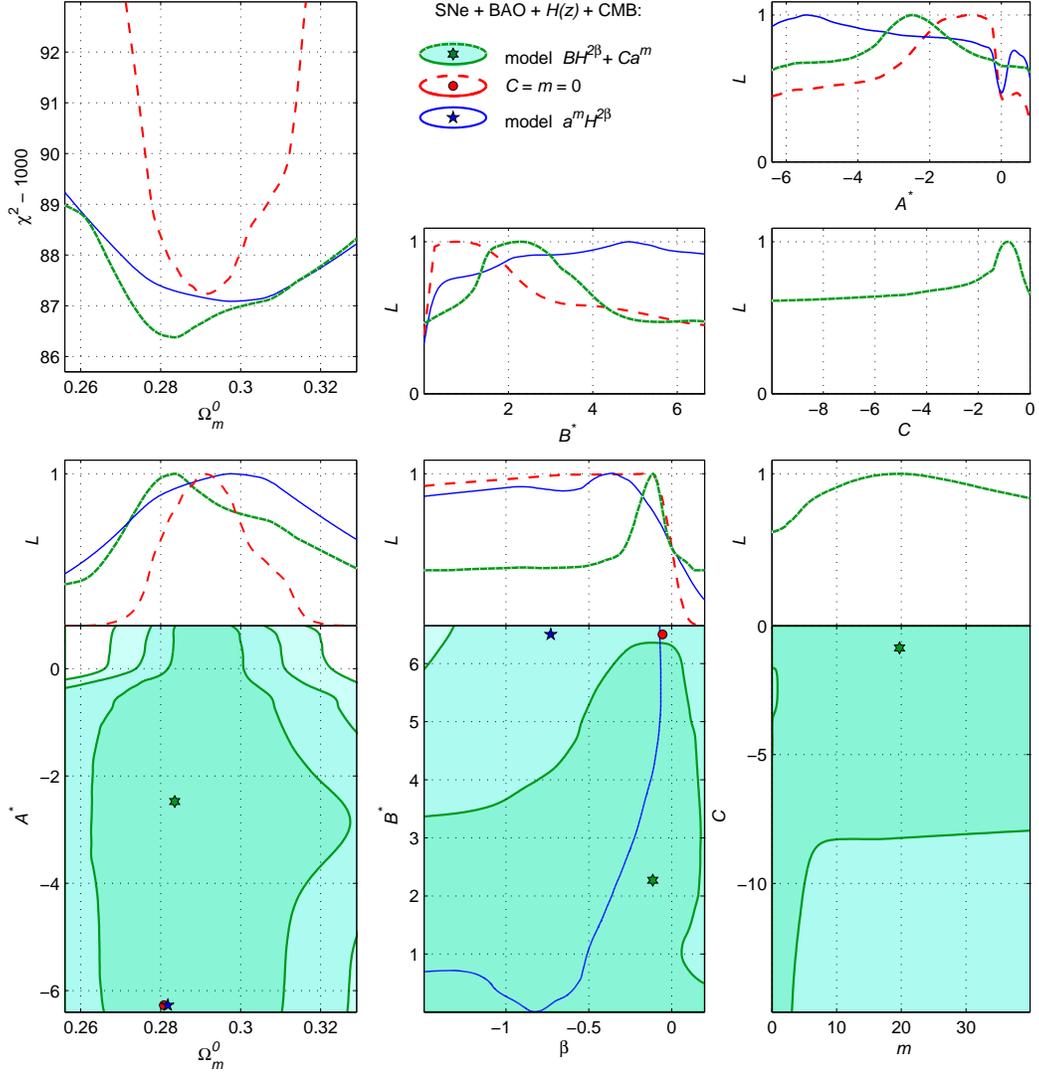}}
\caption{Contours plots, likelihoods and $\chi^2_{\mathrm{tot}}(\Omega_m^0)$ for the
model (\ref{Eqcont1}).  The cases $C=0$ (red) and the model (\ref{Eqcont}) (blue) are
also included for comparison.} \label{F2}
\end{figure}

The second model (\ref{Eqcont1}) contains the terms $B(H/H_0)^{2\beta}$ and
$Ca^m$ in its EoS.  Fig.~\ref{F2} illustrates how this model fits the above datasets, where the contour plots
and the corresponding likelihoods are shown.
This model provides the lowest value $\min\chi^2_{\mathrm{tot}}$ in comparison to the other models,
but contains one extra parameter $C$.
The best fitted values for  $|A|$ and $B$ are not too large and bounded within 1-$\sigma$ area.
Here we compare this model to the case $C=m=0$  and to the model (\ref{Eqcont})
(plotting just $1\sigma$ regions  for the this model in Fig.~\ref{F2}).

The model (\ref{Eqcont1}) provides the best fits in terms of
$ \min\chi^2_{\mathrm{tot}}=1084.92$ in comparison to the other
models, as shown in Table~\ref{Estim} but at the price of increasing the number of free parameters,
since model (\ref{Eqcont1}) has the largest number of
free parameters $N_p=8$ if we include nuisance $H_0$ and $\Omega_b^0$.
Due to this reason, the information criteria, in particular the Akaike information criterion:
\cite{Akaike74}
\[
AIC = \min\chi^2_\Sigma +2N_p
\]
will favour other simpler models with a shorter number of free parameters.

As the  number of free parameters $N_p$ for the model (\ref{Eqcont1}) is too large,
Fig.~\ref{F2} is limited to the panels: $\Omega_m^0- A^*$, $\beta-B^*$,
$C-m$. The corresponding minimums for the $\chi^2_{\mathrm{tot}}$ function are labeled in the plots for each case.
These planes help us to determine the likelihoods for the six free parameters. In
particular, the likelihood $ \mathcal{L}(\Omega_m^0)$ for both models (\ref{Eqcont}) and
(\ref{Eqcont1}) shows the difference in the best fits of $\Omega_m^0$ with larger value
$\Omega_m^0=0.298_{-0.029}^{+0.029}$ and  larger $1\sigma$ range for the models
(\ref{Eqcont}). The best fits for the other parameters in Table~\ref{Estim} also differ
 for both models, the most essential difference is in $A^*$ $B^*$ and $m$.


\section{Conclusions}
\label{Sec8}

In the present paper, a way of constructing a general EoS departing from the FLRW equations is presented.
By assuming the corresponding dependence of a particular fluid in terms of powers of the scale factor,
which at the end is usual in perfect fluids with constant EoS, and powers of the Hubble parameter,
one can easily analyze the early and late universe, which give some information about
the asymptotic behaviors of the cosmological expansion.
Moreover, depending on the dependence of the energy density, future (past) singularities might occur.
While future singularities are well known to occur for some particular EoS parameters,
for instance $w<-1$ leads to a future Big Rip as the energy density increases with the scale factor,
we show also that depending on the EoS, the initial singularity might be a Big Bang-like singularity or a sudden singularity.
In addition, such reconstruction of the EoS and the analysis of the cosmological evolution give rise naturally
to the construction of the EoS for early dark energy, just by imposing on the energy density to become relevant
just before the recombination epoch while decays rapidly after and plays no role at early universe.
This provides a simple way for constructing the effective EoS for early dark energy that can lead to a better knowledge
about the mechanism for alleviating the Hubble tension.
Moreover, here we also present two models for the EoS that can accomplish the behavior for early dark energy
and dark energy itself, as each term might provide a dominance at different periods
of the cosmological evolution depending on the values of the free parameters, modelling the early dark energy and dark energy under the same EoS.

Then, we have tested the models (\ref{O1}) and  (\ref{EoS1}) with observational data, which includes
the Pantheon SNe Ia data, BAO data,  $H(z)$ estimations
and CMB data. The results are shown in Table~\ref{Estim}, which provides good fits at least in terms of
the maximum of the likelihood function, despite the large number of parameters of the models.
Due to this reason the Akaike or Bayesian information criteria \cite{Akaike74,Schwarz78} gives
some preferences to the simplest models with lowest number of parameters, in particular to $w$CDM
and $\Lambda$CDM models, which are particular cases of the models (\ref{O1}) and (\ref{EoS1}).
However, the minimum  $\chi^2_{\mathrm{tot}}$ for the models (\ref{O1}) and (\ref{EoS1}) show
a better fit than such models. From this point of view the model (\ref{O1}) shows the best fit in comparison to the other cases.
 From the best fit values included in Table~\ref{Estim}, and the corresponding EoS given in (\ref{O1}),
we can conclude that the model behaves as a phantom fluid at late universe, while a Big Bang singularity occurs
as the initial state, since the energy density decays at early universe and  radiation and (later on) pressureless matter dominate.
By the results, the power of the Hubble parameter in the EoS (\ref{O1}) can dominate at some intermediate period,
which might be identified to early dark energy.
For the model (\ref{EoS1})  the minimum $\chi^2_{\mathrm{tot}}$ is larger than the other case, and also leads
to larger values for $|A|$ and $B$. In this case, the behavior at late-times is also a phantom-like fluid,
while in the early universe a singularity occurs but its nature is not conclusive,
as the EoS depends on a negative power of the scale factor, such that a type II singularity
might be the initial singularity in this case.

Hence, we have shown a way for constructing not only dark energy EoS but also the so-called early dark energy EoS
that can reveal some properties on the nature of these fluids and  shed some light on issue of the Hubble tension.

\begin{acknowledgments}
This work is supported by the JSPS Grant-in-Aid for Scientific Research (C)
No. 18K03615 (S.N.). The work by SDO was supported by MINECO (Spain), project PID2019-104397GB-I00. DS-CG is funded by the University of Valladolid.
\end{acknowledgments}


\begin{thebibliography}{99}


\bibitem{Bamba:2012cp}
K.~Bamba, S.~Capozziello, S.~Nojiri and S.~D.~Odintsov,
Astrophys. Space Sci. \textbf{342} (2012), 155-228
doi:10.1007/s10509-012-1181-8
[arXiv:1205.3421 [gr-qc]].

\bibitem{Huterer:2017buf}
D.~Huterer and D.~L.~Shafer,
Rept. Prog. Phys. \textbf{81} (2018) no.1, 016901
doi:10.1088/1361-6633/aa997e
[arXiv:1709.01091 [astro-ph.CO]].

\bibitem{Aghanim:2018eyx}
N.~Aghanim \textit{et al.} [Planck],
Astron. Astrophys. \textbf{641} (2020), A6
doi:10.1051/0004-6361/201833910
[arXiv:1807.06209 [astro-ph.CO]].

\bibitem{Aghanim:2018eyx}
N.~Aghanim \textit{et al.} [Planck],
Astron. Astrophys. \textbf{641} (2020), A6
doi:10.1051/0004-6361/201833910
[arXiv:1807.06209 [astro-ph.CO]].

\bibitem{Riess:2019cxk}
A.~G.~Riess, S.~Casertano, W.~Yuan, L.~M.~Macri and D.~Scolnic,
Astrophys. J. \textbf{876} (2019) no.1, 85
doi:10.3847/1538-4357/ab1422
[arXiv:1903.07603 [astro-ph.CO]].

\bibitem{Verde:2019ivm}
L.~Verde, T.~Treu and A.~G.~Riess,
Nature Astron. \textbf{3}, 891
doi:10.1038/s41550-019-0902-0
[arXiv:1907.10625 [astro-ph.CO]].

\bibitem{DiValentino:2021izs}
E.~Di Valentino, O.~Mena, S.~Pan, L.~Visinelli, W.~Yang, A.~Melchiorri, D.~F.~Mota, A.~G.~Riess and J.~Silk,
[arXiv:2103.01183 [astro-ph.CO]].

\bibitem{DiValentino:2020zio}
E.~Di Valentino, L.~A.~Anchordoqui, O.~Akarsu, Y.~Ali-Haimoud, L.~Amendola, N.~Arendse, M.~Asgari, M.~Ballardini, S.~Basilakos and E.~Battistelli, \textit{et al.}
[arXiv:2008.11284 [astro-ph.CO]].

\bibitem{Freedman:2020dne}
W.~L.~Freedman, B.~F.~Madore, T.~Hoyt, I.~S.~Jang, R.~Beaton, M.~G.~Lee, A.~Monson, J.~Neeley and J.~Rich,
doi:10.3847/1538-4357/ab7339
[arXiv:2002.01550 [astro-ph.GA]].

\bibitem{Wong:2019kwg}
K.~C.~Wong, S.~H.~Suyu, G.~C.~F.~Chen, C.~E.~Rusu, M.~Millon, D.~Sluse, V.~Bonvin, C.~D.~Fassnacht, S.~Taubenberger and M.~W.~Auger, \textit{et al.}
Mon. Not. Roy. Astron. Soc. \textbf{498} (2020) no.1, 1420-1439
doi:10.1093/mnras/stz3094
[arXiv:1907.04869 [astro-ph.CO]].

\bibitem{Pesce:2020xfe}
D.~W.~Pesce, J.~A.~Braatz, M.~J.~Reid, A.~G.~Riess, D.~Scolnic, J.~J.~Condon, F.~Gao, C.~Henkel, C.~M.~V.~Impellizzeri and C.~Y.~Kuo, \textit{et al.}
Astrophys. J. Lett. \textbf{891} (2020) no.1, L1
doi:10.3847/2041-8213/ab75f0
[arXiv:2001.09213 [astro-ph.CO]].

\bibitem{Dainotti:2021pqg}
M.~G.~Dainotti, B.~De Simone, T.~Schiavone, G.~Montani, E.~Rinaldi and G.~Lambiase,
[arXiv:2103.02117 [astro-ph.CO]].

\bibitem{Renzi:2020fnx}
F.~Renzi and A.~Silvestri,
[arXiv:2011.10559 [astro-ph.CO]].


\bibitem{Poulin:2018cxd}
V.~Poulin, T.~L.~Smith, T.~Karwal and M.~Kamionkowski,
Phys. Rev. Lett. \textbf{122} (2019) no.22, 221301
doi:10.1103/PhysRevLett.122.221301
[arXiv:1811.04083 [astro-ph.CO]].

\bibitem{Mortsell:2018mfj}
E.~M\"ortsell and S.~Dhawan,
JCAP \textbf{09} (2018), 025
doi:10.1088/1475-7516/2018/09/025
[arXiv:1801.07260 [astro-ph.CO]].

\bibitem{Niedermann:2020dwg}
F.~Niedermann and M.~S.~Sloth,
Phys. Rev. D \textbf{102} (2020) no.6, 063527
doi:10.1103/PhysRevD.102.063527
[arXiv:2006.06686 [astro-ph.CO]].

\bibitem{Garcia:2020sjl}
L.~\'A.~Garc\'\i{}a, L.~Casta\~neda and J.~M.~Tejeiro,
New Astron. \textbf{84}, 101503 (2021)
doi:10.1016/j.newast.2020.101503
[arXiv:2009.07357 [astro-ph.CO]].

\bibitem{Ye:2020btb}
G.~Ye and Y.~S.~Piao,
Phys. Rev. D \textbf{101}, no.8, 083507 (2020)
doi:10.1103/PhysRevD.101.083507
[arXiv:2001.02451 [astro-ph.CO]].

\bibitem{DiValentino:2017iww}
E.~Di Valentino, A.~Melchiorri and O.~Mena,
Phys. Rev. D \textbf{96} (2017) no.4, 043503
doi:10.1103/PhysRevD.96.043503
[arXiv:1704.08342 [astro-ph.CO]].

\bibitem{Yang:2018euj}
W.~Yang, S.~Pan, E.~Di Valentino, R.~C.~Nunes, S.~Vagnozzi and D.~F.~Mota,
JCAP \textbf{09} (2018), 019
doi:10.1088/1475-7516/2018/09/019
[arXiv:1805.08252 [astro-ph.CO]].

\bibitem{Pan:2019jqh}
S.~Pan, W.~Yang, C.~Singha and E.~N.~Saridakis,
Phys. Rev. D \textbf{100} (2019) no.8, 083539
doi:10.1103/PhysRevD.100.083539
[arXiv:1903.10969 [astro-ph.CO]].

\bibitem{Gomez-Valent:2020mqn}
A.~G\'omez-Valent, V.~Pettorino and L.~Amendola,
Phys. Rev. D \textbf{101} (2020) no.12, 123513
doi:10.1103/PhysRevD.101.123513
[arXiv:2004.00610 [astro-ph.CO]].

\bibitem{Pan:2020zza}
S.~Pan, G.~S.~Sharov and W.~Yang,
Phys. Rev. D \textbf{101} (2020) no.10, 103533
doi:10.1103/PhysRevD.101.103533
[arXiv:2001.03120 [astro-ph.CO]].

\bibitem{DEramo:2018vss}
F.~D'Eramo, R.~Z.~Ferreira, A.~Notari and J.~L.~Bernal,
JCAP \textbf{11} (2018), 014
doi:10.1088/1475-7516/2018/11/014
[arXiv:1808.07430 [hep-ph]].

\bibitem{Vagnozzi:2019ezj}
S.~Vagnozzi,
Phys. Rev. D \textbf{102} (2020) no.2, 023518
doi:10.1103/PhysRevD.102.023518
[arXiv:1907.07569 [astro-ph.CO]].

\bibitem{Nunes:2018xbm}
R.~C.~Nunes,
JCAP \textbf{05} (2018), 052
doi:10.1088/1475-7516/2018/05/052
[arXiv:1802.02281 [gr-qc]].

\bibitem{Wang:2020zfv}
D.~Wang and D.~Mota,
Phys. Rev. D \textbf{102} (2020) no.6, 063530
doi:10.1103/PhysRevD.102.063530
[arXiv:2003.10095 [astro-ph.CO]].

\bibitem{Odintsov:2020qzd}
S.~D.~Odintsov, D.~S.~C.~G\'omez and G.~S.~Sharov,
[arXiv:2011.03957 [gr-qc]].

\bibitem{Braglia:2020auw}
M.~Braglia, M.~Ballardini, F.~Finelli and K.~Koyama,
Phys. Rev. D \textbf{103}, no.4, 043528 (2021)
doi:10.1103/PhysRevD.103.043528
[arXiv:2011.12934 [astro-ph.CO]].

\bibitem{Huterer:2000mj}
D.~Huterer and M.~S.~Turner,
Phys. Rev. D \textbf{64} (2001), 123527
doi:10.1103/PhysRevD.64.123527
[arXiv:astro-ph/0012510 [astro-ph]].

\bibitem{Nojiri:2005sr}
S.~Nojiri and S.~D.~Odintsov,
Phys. Rev. D \textbf{72} (2005), 023003
doi:10.1103/PhysRevD.72.023003
[arXiv:hep-th/0505215 [hep-th]].

\bibitem{Capozziello:2005pa}
S.~Capozziello, V.~F.~Cardone, E.~Elizalde, S.~Nojiri and S.~D.~Odintsov,
Phys. Rev. D \textbf{73} (2006), 043512
doi:10.1103/PhysRevD.73.043512
[arXiv:astro-ph/0508350 [astro-ph]].

\bibitem{Gerardi:2019obr}
F.~Gerardi, M.~Martinelli and A.~Silvestri,
JCAP \textbf{07}, 042 (2019)
doi:10.1088/1475-7516/2019/07/042
[arXiv:1902.09423 [astro-ph.CO]].

\bibitem{Brevik:2017msy}
I.~Brevik, \O{}.~Gr\o{}n, J.~de Haro, S.~D.~Odintsov and E.~N.~Saridakis,
Int. J. Mod. Phys. D \textbf{26} (2017) no.14, 1730024
doi:10.1142/S0218271817300245
[arXiv:1706.02543 [gr-qc]].

\bibitem{Cataldo:2005qh}
M.~Cataldo, N.~Cruz and S.~Lepe,
Phys. Lett. B \textbf{619} (2005), 5-10
doi:10.1016/j.physletb.2005.05.029
[arXiv:hep-th/0506153 [hep-th]].

\bibitem{Cruz:2018yrr}
N.~Cruz, E.~Gonz\'alez, S.~Lepe and D.~S\'aez-Chill\'on G\'omez,
JCAP \textbf{12} (2018), 017
doi:10.1088/1475-7516/2018/12/017
[arXiv:1807.10729 [gr-qc]].


\bibitem{Elizalde:2008yf}
E.~Elizalde, S.~Nojiri, S.~D.~Odintsov, D.~Saez-Gomez and V.~Faraoni,
Phys. Rev. D \textbf{77}, 106005 (2008)
doi:10.1103/PhysRevD.77.106005
[arXiv:0803.1311 [hep-th]].

\bibitem{Leanizbarrutia:2014xta}
I.~Leanizbarrutia and D.~S\'aez-G\'omez,
Phys. Rev. D \textbf{90} (2014) no.6, 063508
doi:10.1103/PhysRevD.90.063508
[arXiv:1404.3665 [astro-ph.CO]].

\bibitem{SaezGomez:2008uj}
D.~Saez-Gomez,
Gen. Rel. Grav. \textbf{41} (2009), 1527-1538
doi:10.1007/s10714-008-0724-3
[arXiv:0809.1311 [hep-th]].

\bibitem{Volovik:2000ua}
G.~E.~Volovik,
Phys. Rept. \textbf{351} (2001), 195-348
doi:10.1016/S0370-1573(00)00139-3
[arXiv:gr-qc/0005091 [gr-qc]].

\bibitem{Rosu:2020tov}
H.~C.~Rosu, S.~C.~Mancas and C.~C.~Hsieh,
[arXiv:2010.01720 [gr-qc]].




\bibitem{Ade:2013zuv}
P.~A.~R.~Ade \textit{et al.} [Planck],
Astron. Astrophys. \textbf{571} (2014), A16
doi:10.1051/0004-6361/201321591
[arXiv:1303.5076 [astro-ph.CO]].

\bibitem{Ade:2015xua}
P.~A.~R.~Ade \textit{et al.} [Planck],
Astron. Astrophys. \textbf{594} (2016), A13
doi:10.1051/0004-6361/201525830
[arXiv:1502.01589 [astro-ph.CO]].


\bibitem{Odintsov:2017qif}
S.~D.~Odintsov, D.~S\'aez-Chill\'on G\'omez and G.~S.~Sharov,
Eur. Phys. J. C \textbf{77} (2017) no.12, 862
doi:10.1140/epjc/s10052-017-5419-z
[arXiv:1709.06800 [gr-qc]].

\bibitem{Odintsov:2018qug}
S.~D.~Odintsov, D.~Saez-Chillon Gomez and G.~S.~Sharov,
Phys. Rev. D \textbf{99} (2019) no.2, 024003
doi:10.1103/PhysRevD.99.024003
[arXiv:1807.02163 [gr-qc]].

\bibitem{Odintsov:2020voa}
S.~D.~Odintsov, D.~Saez-Chillon Gomez and G.~S.~Sharov,
Phys. Rev. D \textbf{101} (2020) no.4, 044010
doi:10.1103/PhysRevD.101.044010
[arXiv:2001.07945 [gr-qc]].

\bibitem{Sharov:2015ifa}
G.~S.~Sharov,
JCAP \textbf{06} (2016), 023
doi:10.1088/1475-7516/2016/06/023
[arXiv:1506.05246 [gr-qc]].

\bibitem{Pan:2016ngu}
S.~Pan and G.~S.~Sharov,
Mon. Not. Roy. Astron. Soc. \textbf{472} (2017) no.4, 4736-4749
doi:10.1093/mnras/stx2278
[arXiv:1609.02287 [gr-qc]].

\bibitem{Akaike74}
H.~Akaike, IEEE Trans. Auto. Control AC-19, 716 (1974).

\bibitem{Schwarz78}
G.~Schwarz, Ann. Statist. 6 (2) 461 (1978).

\bibitem{Scolnic:2017caz}
D.~M.~Scolnic, D.~O.~Jones, A.~Rest, Y.~C.~Pan, R.~Chornock, R.~J.~Foley, M.~E.~Huber, R.~Kessler, G.~Narayan and A.~G.~Riess, \textit{et al.}
Astrophys. J. \textbf{859} (2018) no.2, 101
doi:10.3847/1538-4357/aab9bb
[arXiv:1710.00845 [astro-ph.CO]].

\bibitem{Eisenstein:2005su}
D.~J.~Eisenstein \textit{et al.} [SDSS],
Astrophys. J. \textbf{633} (2005), 560-574
doi:10.1086/466512
[arXiv:astro-ph/0501171 [astro-ph]].

\bibitem{Aylor:2019} 
K.~Aylor, M.~Joy, L.~Knox, M.~Millea, S.~Raghunathan and W.~L.~K.~Wu,
Astrophys. J. \textbf{874}, no.1, 4 (2019)
doi:10.3847/1538-4357/ab0898
[arXiv:1811.00537 [astro-ph.CO]];
L.~Knox and M.~Millea,
Phys. Rev. D \textbf{101}, no.4, 043533 (2020)
doi:10.1103/PhysRevD.101.043533
[arXiv:1908.03663 [astro-ph.CO]].




\bibitem{Percival:2009xn}
W.~J.~Percival \textit{et al.} [SDSS],
Mon. Not. Roy. Astron. Soc. \textbf{401} (2010), 2148-2168
doi:10.1111/j.1365-2966.2009.15812.x
[arXiv:0907.1660 [astro-ph.CO]].

\bibitem{Beutler:2011hx}
F.~Beutler, C.~Blake, M.~Colless, D.~H.~Jones, L.~Staveley-Smith, L.~Campbell, Q.~Parker, W.~Saunders and F.~Watson,
Mon. Not. Roy. Astron. Soc. \textbf{416} (2011), 3017-3032
doi:10.1111/j.1365-2966.2011.19250.x
[arXiv:1106.3366 [astro-ph.CO]].

\bibitem{Blake:2011en}
C.~Blake, E.~Kazin, F.~Beutler, T.~Davis, D.~Parkinson, S.~Brough, M.~Colless, C.~Contreras, W.~Couch and S.~Croom, \textit{et al.}
Mon. Not. Roy. Astron. Soc. \textbf{418} (2011), 1707-1724
doi:10.1111/j.1365-2966.2011.19592.x
[arXiv:1108.2635 [astro-ph.CO]].

\bibitem{Padmanabhan:2012hf}
N.~Padmanabhan, X.~Xu, D.~J.~Eisenstein, R.~Scalzo, A.~J.~Cuesta, K.~T.~Mehta and E.~Kazin,
Mon. Not. Roy. Astron. Soc. \textbf{427} (2012) no.3, 2132-2145
doi:10.1111/j.1365-2966.2012.21888.x
[arXiv:1202.0090 [astro-ph.CO]].

\bibitem{Chuang:2012qt}
C.~H.~Chuang and Y.~Wang,
Mon. Not. Roy. Astron. Soc. \textbf{435} (2013), 255-262
doi:10.1093/mnras/stt1290
[arXiv:1209.0210 [astro-ph.CO]].

\bibitem{Chuang:2013hya}
C.~H.~Chuang, F.~Prada, A.~J.~Cuesta, D.~J.~Eisenstein, E.~Kazin, N.~Padmanabhan, A.~G.~Sanchez, X.~Xu, F.~Beutler and M.~Manera, \textit{et al.}
Mon. Not. Roy. Astron. Soc. \textbf{433} (2013), 3559
doi:10.1093/mnras/stt988
[arXiv:1303.4486 [astro-ph.CO]].

\bibitem{Ross:2014qpa}
A.~J.~Ross, L.~Samushia, C.~Howlett, W.~J.~Percival, A.~Burden and M.~Manera,
Mon. Not. Roy. Astron. Soc. \textbf{449} (2015) no.1, 835-847
doi:10.1093/mnras/stv154
[arXiv:1409.3242 [astro-ph.CO]].

\bibitem{Anderson:2013zyy}
L.~Anderson \textit{et al.} [BOSS],
Mon. Not. Roy. Astron. Soc. \textbf{441} (2014) no.1, 24-62
doi:10.1093/mnras/stu523
[arXiv:1312.4877 [astro-ph.CO]].

\bibitem{Oka:2013cba}
A.~Oka, S.~Saito, T.~Nishimichi, A.~Taruya and K.~Yamamoto,
Mon. Not. Roy. Astron. Soc. \textbf{439} (2014), 2515-2530
doi:10.1093/mnras/stu111
[arXiv:1310.2820 [astro-ph.CO]].

\bibitem{Font-Ribera:2013wce}
A.~Font-Ribera \textit{et al.} [BOSS],
JCAP \textbf{05} (2014), 027
doi:10.1088/1475-7516/2014/05/027
[arXiv:1311.1767 [astro-ph.CO]].

\bibitem{Delubac:2014aqe}
T.~Delubac \textit{et al.} [BOSS],
Astron. Astrophys. \textbf{574} (2015), A59
doi:10.1051/0004-6361/201423969
[arXiv:1404.1801 [astro-ph.CO]].





\bibitem{Simon:2004tf}
J.~Simon, L.~Verde and R.~Jimenez,
Phys. Rev. D \textbf{71} (2005), 123001
doi:10.1103/PhysRevD.71.123001
[arXiv:astro-ph/0412269 [astro-ph]].

\bibitem{Stern:2009ep}
D.~Stern, R.~Jimenez, L.~Verde, M.~Kamionkowski and S.~A.~Stanford,
JCAP \textbf{02} (2010), 008
doi:10.1088/1475-7516/2010/02/008
[arXiv:0907.3149 [astro-ph.CO]].

\bibitem{Moresco:2012jh}
M.~Moresco, A.~Cimatti, R.~Jimenez, L.~Pozzetti, G.~Zamorani, M.~Bolzonella, J.~Dunlop, F.~Lamareille, M.~Mignoli and H.~Pearce, \textit{et al.}
JCAP \textbf{08} (2012), 006
doi:10.1088/1475-7516/2012/08/006
[arXiv:1201.3609 [astro-ph.CO]].

\bibitem{Zhang:2012mp}
C.~Zhang, H.~Zhang, S.~Yuan, T.~J.~Zhang and Y.~C.~Sun,
Res. Astron. Astrophys. \textbf{14} (2014) no.10, 1221-1233
doi:10.1088/1674-4527/14/10/002
[arXiv:1207.4541 [astro-ph.CO]].

\bibitem{Moresco:2015cya}
M.~Moresco,
Mon. Not. Roy. Astron. Soc. \textbf{450} (2015) no.1, L16-L20
doi:10.1093/mnrasl/slv037
[arXiv:1503.01116 [astro-ph.CO]].

\bibitem{Moresco:2016mzx}
M.~Moresco, L.~Pozzetti, A.~Cimatti, R.~Jimenez, C.~Maraston, L.~Verde, D.~Thomas, A.~Citro, R.~Tojeiro and D.~Wilkinson,
JCAP \textbf{05} (2016), 014
doi:10.1088/1475-7516/2016/05/014
[arXiv:1601.01701 [astro-ph.CO]].

\bibitem{Ratsimbazafy:2017vga}
A.~L.~Ratsimbazafy, S.~I.~Loubser, S.~M.~Crawford, C.~M.~Cress, B.~A.~Bassett, R.~C.~Nichol and P.~V\"ais\"anen,
Mon. Not. Roy. Astron. Soc. \textbf{467} (2017) no.3, 3239-3254
doi:10.1093/mnras/stx301
[arXiv:1702.00418 [astro-ph.CO]].


\bibitem{Chen:2018dbv}
L.~Chen, Q.~G.~Huang and K.~Wang,
JCAP \textbf{02} (2019), 028
doi:10.1088/1475-7516/2019/02/028
[arXiv:1808.05724 [astro-ph.CO]].

\bibitem{HuSugiyama95}
 W. Hu and N. Sugiyama,
  Astrophys. J. \textbf{471} (1996), 542-570, [arXiv:astro-ph/9510117 [astro-ph]].

 \end{thebibliography}
\end{document}